\documentclass[superscriptaddress,amsmath,amssymb,aps,pre,twocolumn]{revtex4-2}

\usepackage{graphicx}% Include figure files
\usepackage{dcolumn}% Align table columns on decimal point
\usepackage{bm}% bold math
\usepackage{hyperref}% add hypertext capabilities
\usepackage{xcolor}
\usepackage[normalem]{ulem}
\usepackage[linesnumbered,ruled,vlined]{algorithm2e}
\usepackage{subcaption}
\usepackage[justification=raggedright,singlelinecheck=false]{caption}

\begin{document}

\preprint{APS/123-QED}

\title{Entropy of Liquids and Glasses from Recurring Structural Patterns}% 

\author{Nina Javerzat}
\email[Corresponding author: ]{nina.javerzat@univ-grenoble-alpes.fr}
\affiliation{Univ. Grenoble Alpes, CNRS, Institut Fourier, 38402 Saint-
Martin-d’H\`eres, France}

\author{Gerhard Jung}
\affiliation{Institut f\"ur Theoretische Physik, Universit\"at Innsbruck, 6020 Innsbruck, Austria}

\author{Jorge Kurchan}
\affiliation{Laboratoire de Physique de l'Ecole Normale Sup\'erieure, ENS, Universit\'e PSL, CNRS, Sorbonne Universit\'e, Universit\'e de Paris, F-75005 Paris, France}

\author{Misaki Ozawa}
\affiliation{Univ. Grenoble Alpes, CNRS, LIPhy, 38000 Grenoble, France}

\date{\today}% It is always \today, today,
             %  but any date may be explicitly specified

\begin{abstract}
We compute the low-temperature configurational entropy of a two-dimensional supercooled liquid. 
Our method, based on a higher-dimensional version of the Grassberger--Procaccia algorithm, can be implemented in a manner that is entirely agnostic with respect to both the dynamics and the theoretical framework, as any genuine notion of order should be. 
In this construction, entropy is obtained as the decay rate of recurrent structural patterns with increasing patch size, directly linking entropy reduction to the growing persistence of amorphous order. 
Because the method requires only particle positions, without any knowledge of the interaction potential or even of the particle sizes, it can be applied directly to both equilibrium and nonequilibrium aging configurations. 
The resulting configurational entropy, together with the higher-order R\'enyi complexities, agree quantitatively with values obtained from conventional definitions. 
Remarkably, the entropies measured during aging coincide with their equilibrium counterparts when compared at the same inherent-structure energy.\end{abstract}

%\keywords{Suggested keywords}%Use showkeys class option if keyword
                              %display desired
\maketitle

%\tableofcontents

%%%%%%%%%%%%%%%%%%%%%%%%%%%%%%%%%%%%%%%%%%%%%%%%%%%%%%%%%%%%%%%%%%%%%%%%%%%%%%%%%%%%%%%%%%%%%%%%%%%%%%%%%%%%%%%%%%
%%%%%%%%%%%%%%%%%%%%%%%%%%%%%%%%%%%%%%%%%%%%%%%%%%%%%%%%%%%%%%%%%%%%%%%%%%%%%%%%%%%%%%%%%%%%%%%%%%%%%%%%%%%%%%%%%%

%\section{Introduction}
\indent\textit{Introduction---}The liquid-to-glass transition occurs without the appearance of any evident long-range structural order. As the temperature decreases, the viscosity of a supercooled liquid increases dramatically, eventually becoming so large that the system falls out of equilibrium on experimental timescales and forms a glass. Yet, at the level of particle configurations, the system still appears structurally disordered, at least naively, much like a liquid~\cite{berthier2011theoretical,binder2011glassy,ediger1996supercooled}. For this reason, considerable effort has been devoted to identifying more subtle forms of order associated with glass formation, such as crystalline orders~\cite{tanaka2010critical,tanaka2019revealing}, medium-range orders~\cite{tah2017glass,kumawat2025growth}, higher-order correlations~\cite{singh2023intermediate}, locally favored structures~\cite{tarjus2005frustration,coslovich2007understanding,royall2015role}, and point-to-set correlations~\cite{biroli2008thermodynamic,yaida2016point,cammarota2012patch}.

What are the characteristics that we expect from a definition of `spatial order'?  The measure of order should depend on the configuration, and  be independent of a particular theoretical framework.
Consider a crystal: for a large sample, we do not need any theory - or indeed knowledge of the dynamics - to tell us that we have crystalline order. A natural and very general way to find such a measure for an arbitrary, non-periodic form of order originates from the question: how much information does one need to transmit to determine the position of all particles with a given precision $\epsilon$?  For a perfect crystal, the answer is simple: we specify the elementary cell, and then instruct the recipient to repeat the cell. But we know other forms of order: for a quasicrystal, one can show there are patches of all sizes that repeat very frequently, so if we look at a perfect quasicrystal the amount of information one needs to transmit is small,  subextensive in the sample size: patch repetition induces low entropy~\cite{kurchan2010order}. The problem becomes slightly more complicated when on top of this order we have some thermal excitations: how shall we disentangle them from the underlying order? One solution to this problem is to collect several snapshots that may be relatively close in time, and use the fact that fast fluctuations (`vibrational degrees of freedom') will average away~\cite{kurchan2010order}. 

%These repeating motifs are affected by thermal noise - in a similar way as the crystals are - and again, we may either consider time averaging to factor this away.

Years ago it was pointed out~\cite{kurchan2010order} that the size and recurrence of repeated patches should increase as the liquid is further supercooled and configurational entropy drops, signaling the growth of amorphous order. Low configurational entropy means that, setting aside fast degrees of freedom and focusing on the disordered structural backbone, there are relatively few possible particle configurations. This can only occur if the system is a {\it collage} of a small number of distinct patches, just as a text has very little information if it contains many stereotyped phrases.
The connection between particle configurations, growing amorphous order, and the entropy crisis associated with the glass transition is thus natural~\cite{kurchan2010order,fraenkel2024information,kurchan2025comptes}.

A very similar problem arose years ago in the theory of dynamical systems: the trajectories have motifs that tend to repeat themselves in time, an example being the case when an orbit `shadows' a periodic orbit for a while and then flies away. Counting the information taking into account these repetitions or quasi-repetitions is, technically speaking, computing the Kolmogorov-Sinai entropy~\cite{Kolmogorov1958,Sinai1959}.
Grassberger and Procaccia found a clever and very practical way of doing this, displaying together in a meaningful way the similarities between pieces of trajectories~\cite{grassberger1983estimation}. The very same method may be adopted here in the glass problem, considering spatial distributions (configurational) rather than temporal ones (trajectories).
The next question, how to disentangle the effect of `vibrations' is very much like how do we recognize motifs in trajectories when they are measured with some noise. In the case of a crystal, when two patch sizes are large enough, their coincidence adds coherently while the thermal noise on them is random: for long lengths we do not need time averaging, in principle, to detect order.

In this paper, we provide a numerical demonstration of the Grassberger-Procaccia method, extending it beyond the level of a phenomenological argument to a quantitative measurement by overcoming several technical challenges with state-of-the-art numerical methods. We study a two-dimensional glass-forming liquid that, according to recent work, can be equilibrated down to extremely low temperature and for which several independent estimates of the configurational entropy via conventional methods are available~\cite{jung2025numerical}. We develop a new algorithm to identify similar patches in amorphous configurations with high precision, which allows us to estimate the Rényi complexity~\cite{javerzat2025renyi} of order $m$ and its $m \to 1$ limit, corresponding to the configurational entropy. Clearly, these entropies may be inferred solely from snapshots, without requiring any additional information such as the interaction potential or even the particle sizes. In spite of this, our entropy estimates are quantitatively consistent with conventional measures based on, for example, Franz-Parisi potentials~\cite{franz1998effective,berthier2014novel,berthier2017configurational}. 

These entropies remain well defined and measurable even out of equilibrium, and we further demonstrate that they decrease during aging. Remarkably, we find that at given inherent structure energy, the entropy is the same whether the system is aging or in equilibrium at a higher temperature. 

Several previous works have explored how entropy can be inferred directly from configurations or snapshots, using network-based methods~\cite{vink2002configurational,brujic2007measuring,zhao2022measuring}, data-compression algorithms~\cite{avinery2019universal,martiniani2019quantifying,zu2020information,douglass2024complexity,fraenkel2024information}, and machine-learning techniques~\cite{nir2020machine,gu2022thermodynamics}. 
Our approach follows a different route, based on the recurrence of amorphous structural patches. 
In this view, the decrease of entropy is interpreted as the growth of amorphous order, encoded in the increasing recurrence of structural patterns. 
Our study therefore reinforces the notion of amorphous order and connects it directly to the entropy associated with glass formation.
\begin{figure}
\includegraphics[width=\linewidth]{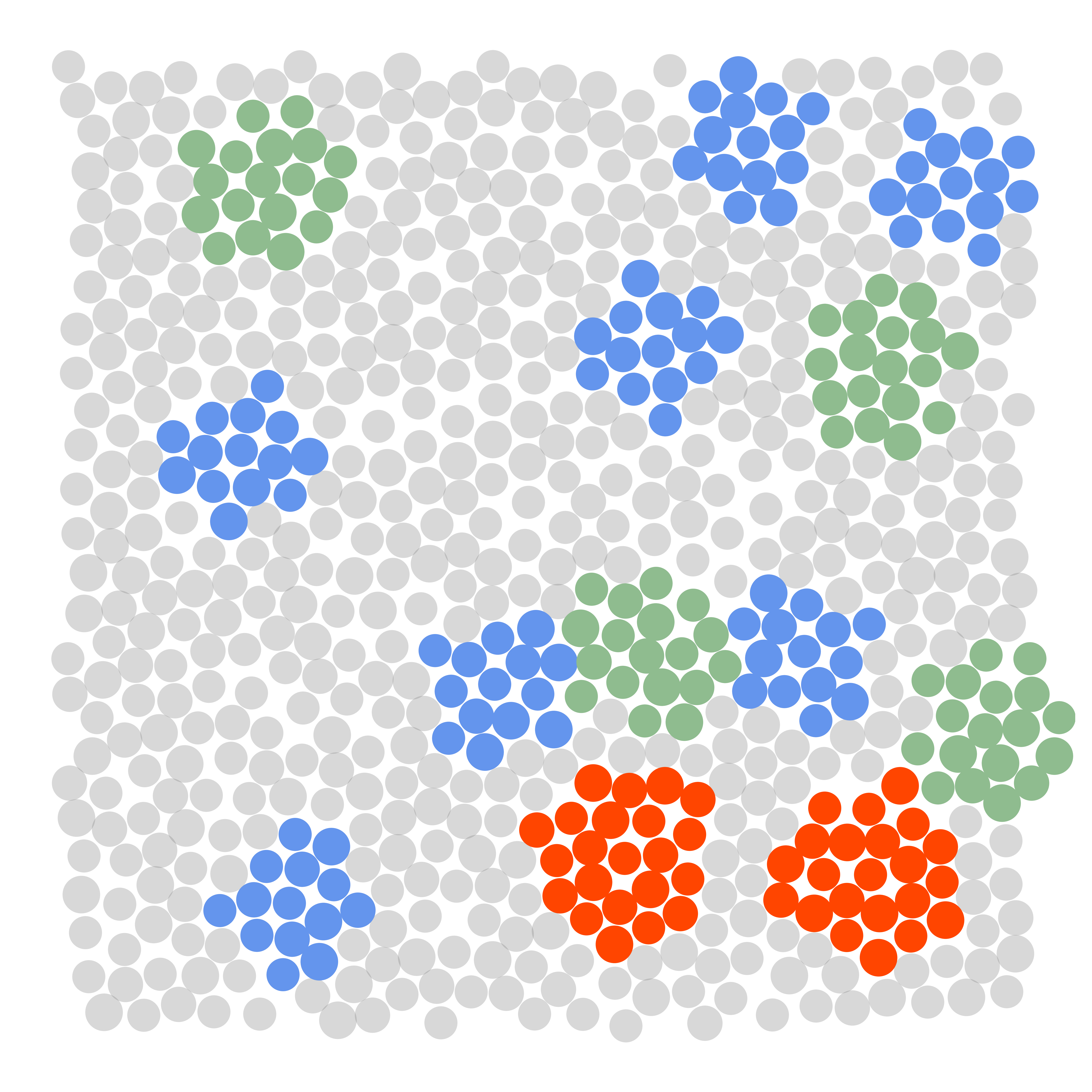}
\caption{Occurrences of similar patches in an $N=704$ configuration. The most similar occurrences, with distance $\Delta \lesssim 0.3$, are highlighted for patch sizes $d=14$, $18$, and $22$ in blue, green, and red, respectively. Overlapping patches are omitted for readability.}
\label{fig:snapshot}
\end{figure}
\newline\newline
\indent\textit{Methods---}In this paper, we consider a modified two-dimensional ternary Kob--Andersen mixture (KA2D), for which recent work~\cite{jung2025numerical} has shown that a combination of advanced Monte Carlo techniques (parallel tempering~\cite{hukushima1996exchange}, swap Monte Carlo~\cite{ninarello2017models}, and population annealing~\cite{hukushima2003population}) allows equilibrium sampling down to extremely low temperature for system sizes up to $N=77$ (see Appendix~\ref{sec:glass_model}). We present results for $N=77$, while finite-size effects are discussed in Appendix~\ref{app:numerical} through a comparison with the larger system size $N=704$.
For this model, and with the computational resources currently available, the extraction of entropy from the Grassberger-Procaccia algorithm can be applied only at sufficiently low temperatures, where amorphous order is well developed. In practice, only the $N=77$ system can be equilibrated down to such low temperatures.

In the Boltzmann view of entropy, the configurational entropy, or complexity, $s^{\rm conf}$, is defined as the logarithm of the number of available metastable states~\cite{charbonneau2023spin}. Equivalently, one may adopt a Shannon perspective on the configurational entropy at low temperature, where the partition function $Z$ is decomposed into contributions from individual metastable states labeled by $\alpha$, as $Z \approx \sum_\alpha Z_\alpha$. One then defines
\begin{equation}
    s^{\rm conf} = - \frac{1}{N}\sum_\alpha \rho_\alpha \ln \rho_\alpha,
    \label{eq:s_conf_def_Shannon}
\end{equation}
where $\rho_\alpha = Z_\alpha / Z$ is the probability of finding the system in the metastable state $\alpha$, and $N$ is the number of particles.

The Rényi entropy is a one-parameter generalization of the Shannon entropy~\cite{renyi1961measures}, with parameter $m$, and has attracted considerable attention in various areas of physics, because of its accessibility in both analytical calculations and experimental measurements~\cite{calabrese2009entanglement,kaufman2016quantum} (see review~\cite{ozawa2024perspective}). The Rényi counterpart of the configurational entropy, the $\mathrm{R\acute{e}nyi}$ complexity~\cite{javerzat2025renyi}, is defined as
\begin{equation}
 s^{\mathrm{R\acute{e}nyi}}_{m} = - \frac{1}{N(m-1)} \ln \sum_\alpha (\rho_\alpha)^m .
 \label{eq:def_Renyi_main}
\end{equation}
One can easily verify that the limit $m \to 1$ recovers the configurational entropy in Eq.~(\ref{eq:s_conf_def_Shannon}).
In general, $s^{\mathrm{R\acute{e}nyi}}_{m}$ is a decreasing function of $m$, so that $s^{\rm conf} \geq s^{\mathrm{R\acute{e}nyi}}_{m}$.  In this paper we consider $m \geq 1$.

The configurational entropy $s^{\rm conf}$ is conventionally computed from the free-energy barrier of the {\it quenched} Franz-Parisi potential~\cite{franz1998effective,berthier2016facets} as $s^{\rm conf} = \beta V^{\rm Quench}$, where $\beta$ is the inverse temperature. Similarly, the R\'enyi complexity is related to the generalized annealed Franz-Parisi potential (difference) $V_m^{\rm Anneal}$ for $m$ clones through~\cite{javerzat2025renyi}
\begin{equation}
s^{\rm R\acute{e}nyi}_{m} = \frac{1}{m-1}\beta V_m^{\rm Anneal}.
\label{eq:renyi_vs_anneal_FP}
\end{equation}
Furthermore, both $s^{\rm conf}$ and $s^{\rm R\acute{e}nyi}_{m}$ have been shown to vanish at the same Kauzmann transition temperature $T_K$ in mean-field spin-glass models with one-step replica-symmetry breaking~\cite{javerzat2025renyi}. Thus, $s^{\rm R\acute{e}nyi}_{m}$ also encodes important thermodynamic aspects of the glass transition.

Measuring $s^{\rm conf}$ and $s^{\rm R\acute{e}nyi}_{m}$ generally requires substantial computational effort, including reliable thermodynamic sampling using advanced numerical techniques such as swap Monte Carlo, parallel tempering, biased-ensemble methods, and related approaches~\cite{berthier2014novel,berthier2017configurational,guiselin2022statistical}.

We instead  estimate the Rényi complexity $s^{\mathrm{R\acute{e}nyi}}_{m}$ by building on the Grassberger-Procaccia algorithm, applying to configurational data~\cite{kurchan2010order}. We first prepare $N_s$ independent equilibrium configurations of a system with $N$ particles at temperature $T$. In principle, a single infinitely large configuration would be enough, but in practice we analyze many independent configurations of finite size.

To extract only the configurational contribution, we thermally average each configuration over short Monte Carlo simulations, which removes thermal vibrational motion. This is in the same spirit as subtracting the vibrational entropy from the total entropy to estimate the configurational entropy~\cite{sciortino1999inherent,karmakar2009growing}. Alternatively, we analyze inherent-structure configurations, obtained by quenching instantaneous configurations to local minima of the energy landscape. We find that both procedures lead to essentially the same entropy estimates. For this reason, we present in the main text the results obtained from inherent structures, while results for thermally averaged configurations are reported in the Appendix~\ref{app:numerical}.

We then define configurational patches. The total number of particles is $\mathcal{N}=N_sN$, labeled by $I=1,2,\dots,\mathcal{N}$. Around each particle $I$, we define a patch of size $d$, made of the central particle and its $d-1$ nearest neighbors. Each patch is therefore identified by its central particle index $I$.

The central quantity in the Grassberger-Procaccia framework~\cite{grassberger1983estimation} is the correlation integral $C_m^{(d)}(\epsilon)$, defined as
\begin{equation}\label{eq:Cm1}
    C_m^{(d)}(\epsilon) = \frac{1}{\mathcal{N}} \sum_{I=1}^{\mathcal{N}} \left[ n_I^{(d)}(\epsilon)\right]^{m-1},
\end{equation}
where $n_I^{(d)}(\epsilon)$ denotes the number of patches of size $d$ that are similar to patch $I$ up to a tolerance threshold $\epsilon$, and is defined by
\begin{equation}\label{eq:ni1}
    n_I^{(d)}(\epsilon)= \sum_{\substack{J=1\\ (J\neq I)}}^{\mathcal{N}} \theta\!\left( \epsilon - \Delta_{IJ}^{(d)} \right) ,
\end{equation}
where $\theta(x)$ is the Heaviside step function and $\Delta_{IJ}^{(d)}$ is the distance between patches $I$ and $J$.

As derived in Appendix~\ref{app:derivation}, the R\'enyi complexity can be estimated from $C_m^{(d)}(\epsilon)$ by taking the limits of vanishing threshold and large patch size:
\begin{equation}\label{eq:sm1}
s^{\mathrm{R\acute{e}nyi}}_{m} \simeq -  \lim_{\epsilon\to0}\,\lim_{d \to \infty} \frac{1}{d(m-1)} \ln C_m^{(d)}(\epsilon).
\end{equation}
The main technical challenges are to define and measure the distance $\Delta_{IJ}^{(d)}$ between two patches of size $d$ in a physically meaningful and numerically tractable way, and to properly evaluate the limits $\epsilon \to 0$ and $d \to \infty$.

In particular, computing $\Delta_{IJ}^{(d)}$ is a nontrivial task, as it requires optimization over translations, rotations and flipping, as well as a combinatorial optimization over particle-pair assignments~\cite{sausset2011characterizing,hallett2018local}. We develop an efficient algorithm that extends a state-of-the-art shape-matching method~\cite{Gunde2021,these_miha}, as detailed in Appendix~\ref{app:shape_matching}.
Figure~\ref{fig:snapshot} illustrates recurring structural patches within a prescribed similarity threshold, as identified by our patch-identification algorithm. Although the configuration is amorphous and disordered, similar patches can be recognized.
\begin{figure}
\includegraphics[width=\linewidth]{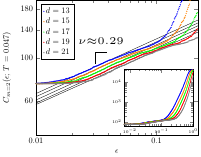}
\caption{Correlation integral $C_m^{(d)}(\epsilon)$ as a function of the distance
threshold $\epsilon$ in the vicinity of the scaling regime, for $m=2$ at
$T=0.047$ and several patch sizes $d$. The straight lines indicate linear
fits in the log--log representation, whose slope gives the exponent $\nu$.
The inset shows $C_m^{(d)}(\epsilon)$ over a broader range of $\epsilon$.}
\label{fig:correlation_integral_main}
\end{figure}
\newline\newline
\indent\textit{Results---}Figure~\ref{fig:correlation_integral_main} (inset) shows $C_m^{(d)}$ as a function of $\epsilon$ for $m=2$ at $T=0.047$, which is well below the empirical (kinetic) glass transition temperature $T_g \approx 0.15$ (see Appendix~\ref{sec:glass_model}). When $\epsilon$ is very large, all $\mathcal{N}$ patches are regarded as identical within this distance threshold, and $C_m^{(d)}$ therefore approaches a constant independent of $d$ (not shown). As $\epsilon$ decreases, for instance below $\epsilon \approx 1$, $C_m^{(d)}$ drops rapidly, reflecting the decrease in the number of identified similar patches. Upon further lowering $\epsilon$, we observe a scaling regime, visible as a linear relation in the log-log plot. Finally, in the limit $\epsilon \to 0$, $C_m^{(d)}$ reaches a plateau due to finite-size effects. This plateau arises because, although the configurations are sampled independently, some of them belong to the same inherent structure at such low temperatures in a finite-size system (see Appendix~\ref{sec:glass_model}). 
To extract the R\'enyi complexity, we focus on the intermediate regime where the scaling behavior is observed, away from the artificial finite-size plateau. Grassberger and Procaccia argued (see e.g., figures in Ref.~\cite{grassberger1983estimation}), and many previous studies of chaotic time-dependent signals have confirmed~\cite{kantz2003nonlinear}, that for small but intermediate values of the threshold $\epsilon$, the correlation integral behaves as
\begin{equation}
C_m^{(d)}(\epsilon) 
\simeq c\, \epsilon^{\nu} 
\exp\!\left[-d(m - 1)s^{\mathrm{R\acute{e}nyi}}_{m}\right].
\label{eq:GP_scaling}
\end{equation}
The correlation integral $C_m^{(d)}(\epsilon)$ increases with $\epsilon$ as $\epsilon^\nu$, where $\nu$ is interpreted as a fractal dimension, see Appendix~\ref{sec:intrinsic_dimension}. Conversely, $C_m^{(d)}(\epsilon)$ decreases as the patch size $d$ increases, and $s^{\mathrm{R\acute{e}nyi}}_{m}$ quantifies this decay rate. Larger values of $s^{\mathrm{R\acute{e}nyi}}_{m}$ indicate a more rapid decay of correlations with patch size, whereas smaller values imply that correlations persist over larger patches.
We indeed confirm that our data for $C_m^{(d)}(\epsilon)$ follow Eq.~(\ref{eq:GP_scaling}), both in their $\epsilon$-dependence and in their $d$-dependence (see Fig.~\ref{fig:correlation_integral_main} and Appendix~\ref{app:numerical}). This expression then provides a practical way to extract $s^{\mathrm{R\acute{e}nyi}}_{m}$ from the asymptotic behavior of $C_m^{(d)}(\epsilon)$ at small $\epsilon$ and large $d$. In this work, we use two distinct methods based on Eq.~(\ref{eq:GP_scaling}) to estimate $s^{\mathrm{R\acute{e}nyi}}_{m}$ from the data for $C_m^{(d)}(\epsilon)$. The two methods yield quantitatively consistent results (see Appendix~\ref{app:numerical}).

In Fig.~\ref{fig:entropies}, we show the estimated $s^{\mathrm{R\acute{e}nyi}}_{m}$ for $m=2$, together with two independent measurements of the same complexity: the annealed Franz-Parisi potential $\beta V_{m=2}^{\rm Anneal}$ and a direct evaluation based on Eq.~(\ref{eq:def_Renyi_main}), in which each metastable state $\alpha$ is approximately identified with an inherent structure~\cite{parmar2023depleting}. We find that $s^{\mathrm{R\acute{e}nyi}}_{m}$ measured from the correlation integral decreases toward the glass transition in quantitative agreement with these two conventional approaches. This is particularly remarkable because our method is inferred solely from configurational snapshots, without any knowledge of the interaction potential and particle sizes.

We then consider the limit $m \to 1$, which corresponds to the configurational entropy, $s^{\rm conf} = \lim_{m \to 1} s_m^{\mathrm{R\acute{e}nyi}}$.
As described in Appendix~\ref{app:derivation}, taking the $m \to 1$ limit of Eq.~(\ref{eq:sm1}) yields
\begin{eqnarray}
    s^{\rm conf} \simeq - \lim_{\epsilon \to 0} \lim_{d \to \infty} \frac{1}{d}
    \ln \tilde C_1^{(d)}(\epsilon),
\end{eqnarray}
where
\begin{equation}\label{eq:Ctilde}
    \ln \tilde C_1^{(d)}(\epsilon)
    = \frac{1}{\mathcal{N}} \sum_{I=1}^{\mathcal{N}} \ln n_I^{(d)}(\epsilon) .
\end{equation}
Thus, the asymptotic behavior of $\ln \tilde C_1^{(d)}(\epsilon)$ in the limits of small $\epsilon$ and large $d$ provides an estimate of $s^{\rm conf}$~\cite{cohen1985computing}. In particular, we expect, and numerically confirm (see Appendix~\ref{app:numerical}), that
\begin{equation}
    \tilde C_1^{(d)}(\epsilon) \simeq c \epsilon^\nu \exp \left[ -d s^{\rm conf} \right] .
    \label{eq:GP_scaling_m1}
\end{equation}
Figure~\ref{fig:entropies} shows the numerically extracted $s^{\rm conf}$, compared with two independent estimates of the configurational entropy: one obtained from the quenched Franz--Parisi potential, and the other from a direct evaluation of Eq.~(\ref{eq:s_conf_def_Shannon}) in which metastable states are approximately identified with inherent structures. We find quantitative agreement among all three estimates. The figure also confirms the theoretical bound $s^{\rm conf} \geq s^{\mathrm{R\acute{e}nyi}}_{m=2}$.
\begin{figure}
\includegraphics[width=\linewidth]{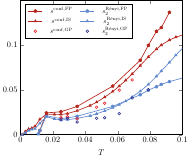}
\caption{
Configurational entropy $s^{\rm conf}$ and R\'enyi complexity
$s^{\mathrm{R\acute{e}nyi}}_{m=2}$, estimated using the
Grassberger--Procaccia (GP) algorithm (open circles). The corresponding estimates
obtained from the free-energy differences of the quenched and annealed
Franz--Parisi (FP) potentials are shown as pentagons, those obtained by
direct enumeration of inherent structures (IS) are shown as triangles.
}
\label{fig:entropies}
\end{figure}
\newline\newline
We wish to note that for $m=2$, the exponent $\nu$ corresponds to the correlation dimension~\cite{Grassberger1983strange}, which measures the intrinsic dimensionality of the manifold effectively sampled by the data~\cite{camastra2016intrinsic}. In our case, although the embedding dimension of a patch is $2d$, we find that the measured values of $\nu$ are much smaller than $2d$, typically of order unity, and decrease systematically upon cooling, as shown in Appendix~\ref{sec:intrinsic_dimension}. This indicates that patch configurations do not explore the full phase space, but instead occupy increasingly constrained and localized regions as $T$ decreases. This behavior is reminiscent of the mean-field picture of glassy phase-space organization, in which low-temperature configurations concentrate into separated metastable regions~\cite{parisi2010mean,krzakala2007gibbs}. Our analysis therefore provides a quantitative finite-dimensional probe of this organization through the intrinsic geometry of recurrent structural patches.
\newline\newline
So far, we have studied the entropy in thermally equilibrium systems. However, our patch  method can be applied directly to out-of-equilibrium systems, since it does not rely on equilibrium thermodynamic sampling, unlike conventional computational approaches.

We therefore measure the entropy of the system along an aging trajectory, as it evolves from higher to lower energies. To this end, we prepare nonequilibrium aging configurations that mimic different stages of aging using population annealing (see Appendix~\ref{sec:glass_model}). Figure~\ref{fig:aging} shows the entropy as a function of energy. We find that the entropy decreases as the energy decreases along the nonequilibrium aging path, consistent with the idea that the system progressively loses access to available metastable states as it explores deeper regions of the energy landscape.

In Fig.~\ref{fig:aging} we see  there is good agreement 
between the entropy of states visited while aging, and equilibrium ones having  the same inherent structure energy \cite{cugliandolo1997energy,grigera1999observation,crisanti2000activated}. This is physically important: it means that along an aging process states are visited as in quasiequilibrium, with an adiabatically decreasing effective temperature - here incorporated through the inherent structure energy.  
\begin{figure}    \includegraphics[width=\linewidth]{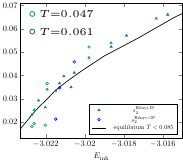}
    \caption{
R\'enyi complexity $s^{\mathrm{R\acute{e}nyi}}_{2}$ measured using the
Grassberger--Procaccia (GP) method (open circles) and from
inherent-structure (IS) statistics (triangles), plotted as a function of
the inherent-structure energy $E_{\mathrm{inh}}$ along out-of-equilibrium
aging paths at target temperatures $T=0.047$ and $T=0.061$.
The black line shows the complexity for the equilibrium data, at temperatures $T<0.085$.
}
    \label{fig:aging}
\end{figure}
%%%%%%%%%%%%%%%%%%%%%%%%%%%%%%%%%%%%%%%%%%%%%%%%%%%%%%%%%%%%%%%%%%%%%%%%%%%%%%%%%%%%%%%%%%%%%%%%%%
%%%%%%%%%%%%%%%%%%%%%%%%%%%%%%%%%%%%%%%%%%%%%%%%%%%%%%%%%%%%%%%%%%%%%%%%%%%%%%%%%%%%%%%%%%%%%%%%%%
\newline\newline
\indent\textit{Conclusion and discussion---}We have numerically estimated the entropies of equilibrium and nonequilibrium configurations of a two-dimensional glass-forming liquid using the Grassberger--Procaccia construction. 
This study provides a proof of concept for the proposed method~\cite{kurchan2010order}. 
For the present model, we have identified a clear scaling regime only at very low temperatures, where amorphous order is well developed. 
This limitation may be system dependent: for example, in models such as silica~\cite{dirindin2025glassy}, where local structural motifs are more pronounced, the scaling regime may be more readily accessible than in Kob--Andersen-type mixtures.
An example where the method could provide physical insight that is difficult to access otherwise is the case of ultrastable glasses~\cite{kapteijns2019fast,leoni2025generating,fan2026ideal,bolton2026ideal,leoni2026computational}. 
Our method can be extended to three-dimensional systems, including molecular systems and systems composed of nonspherical particles, for which patch-identification algorithms are readily available. 
It can also be applied, in principle, to experimental systems such as colloidal suspensions~\cite{brujic2007measuring}, where particle positions can be directly tracked.

In addition, our approach incorporates time-averaged snapshots. 
Such a procedure is experimentally feasible: time-separated scanning tunneling microscopy images of disordered systems can be overlaid, enabling quantitative comparisons and even the tracking of evolving configurations; see, e.g., Ref.~\cite{swartzentruber1996direct}. 
Even when time averaging is not possible, instantaneous configurations with sufficiently large statistics and large patches may allow access to the scaling regime, as tested in Appendix~\ref{app:numerical}.

Finally, the Grassberger--Procaccia construction offers a clear view of how entropy is connected to structural, or amorphous, order. 
Through Eqs.~\eqref{eq:GP_scaling} and~\eqref{eq:GP_scaling_m1},
the entropies are obtained from the decay, with increasing patch size,
of the probability of finding recurrent structural patterns. Smaller entropies therefore imply  patch recurrence of larger patches, or equivalently the persistence of structural correlations over larger length scales. 
Thus, the connection between entropy and amorphous order follows directly from the information theoretic view, without relying on an additional physical scenario linking entropy and length scales, such as the Adam--Gibbs argument~\cite{adam1965temperature}.
%More generally, a procedure that could greatly increase our understanding of real glasses is to compare the results at the end of two different processes  extending our procedure to computing cross-correlations where a patch on a sample (e.g out of equilibrium aging) is compared to patches on another (e.g. equilibrium at a given temperature). \NJ{this sentence is not super clear}
%It would also be intriguing to apply the method to nonequilibrium self-propelled systems, such as dense active matter and biological tissues, where locally disordered packings may nevertheless display recurring patches. In such intrinsically nonequilibrium systems, our method could provide a way to estimate entropies directly from structural data, potentially revealing hidden order within apparently disordered configurations~\cite{martiniani2019quantifying}.
\newline\newline
\indent\textit{Acknowledgements---}We thank Miha Gunde and Nicolas Salles for very helpful discussions on the IRA algorithm. We also thank Ludovic Berthier, Eric Bertin, Romain Mari, and Itamar Procaccia for insightful exchanges.
MO thanks the support by MIAI@Grenoble Alpes and the Agence Nationale de la Recherche under France 2030 with the reference ANR-23-IACL-0006). GJ thanks the Austrian Science Fund (FWF) 10.55776/PAT1139125.
\newline\newline
\indent\textit{Data Availability---}The source codes used in this paper are openly available at \cite{Gitcode}.
\bibliography{reference}
%%%%%%%%%%%%%%%%%%%%%%%%%%%%%%%%%%%%%%%%%%%%%%%%%%%%%%%%%%%%%%%%%%%%%%%%%%%%%%%%%%%%%%%%%%%%%%%%%%%%%%%%%%%%%%%%%%
%%%%%%%%%%%%%%%%%%%%%%%%%%%%%%%%%%%%%%%%%%%%%%%%%%%%%%%%%%%%%%%%%%%%%%%%%%%%%%%%%%%%%%%%%%%%%%%%%%%%%%%%%%%%%%%%%%
%%%%%%%%%%%%%%%%%%%%%%%%%%%%%%%%%%%%%%%%%%%%%%%%%%%%%%%%%%%%%%%%%%%%%%%%%%%%%%%%%%%%%%%%%%%%%%%%%%%%%%%%%%%%%%%%%%

\clearpage

\appendix

\section{Glass model, equilibration and configurational entropy}\label{sec:glass_model}

In this Appendix, we describe the simulation model and the methodology used to generate equilibrium configurations, their thermally averaged counterparts, and the corresponding inherent structures for the Grassberger-Procaccia algorithm. We also explain the conventional procedure used to compute the entropy. Both the methods and the data are identical to those analyzed in Ref.~\cite{jung2025numerical}.

\subsection{Simulation model}

The atomistic glass-forming model studied in this work is closely based on the system introduced in Ref.~\cite{jung2023predicting}. It consists of a modified two-dimensional ternary Kob--Andersen mixture (KA2D), in which particles interact through a Lennard--Jones potential given by
\begin{equation}\label{eq:LJ}
V_{\alpha \beta}(r_{ij}) = \begin{cases}
4 \epsilon_{\alpha \beta} \left[\left(\frac{\sigma_{\alpha \beta}}{r_{ij}}\right)^{12} - \left(\frac{\sigma_{\alpha \beta}}{r_{ij}}\right)^6 + C_0 \right. & \\  \,\left. + C_2\left( \frac{r_{ij}}{\sigma_{\alpha \beta}}\right)^{2}+ C_4\left( \frac{r_{ij}}{\sigma_{\alpha \beta}}\right)^{4}  \right] & r_{ij}<r^\text{cut}_{\alpha \beta}\\
0 & \text{otherwise},
\end{cases}
\nonumber
\end{equation}
where the interparticle distance is defined as $r_{ij} = |\mathbf{r}_i - \mathbf{r}_j|$, with $\mathbf{r}_i$ denoting the position of particle $i$.

The ternary KA2D mixture contains three particle species $\alpha,\beta = {1,2,3}$ with a composition ratio of 5:3:3. Species 1 and 2 interact through the standard non-additive Kob–Andersen parameters, $\epsilon_{11}=1.0$, $\epsilon_{12}=1.5$, $\epsilon_{22}=0.5$ and $\sigma_{11}=1.0$, $\sigma_{12}=0.8$, $\sigma_{22}=0.88.$ In addition, we incorporate a third species with interaction parameters $\epsilon_{13}=0.75$, $\epsilon_{23}=1.5$, $\epsilon_{33}=0.75$ and $\sigma_{13}=0.9$, $\sigma_{23}=0.8$, $\sigma_{33}=0.94$~\cite{parmar2020ultrastable}. The cutoff distance depends on the particle pair and is defined as $r^\text{cut}_{\alpha \beta} = 2.5\sigma_{\alpha \beta}.$ The constants $C_0= 0.04049023795$, $C_2= -0.00970155098$, and $C_4= 0.00062012616$ are chosen such that the potential and its first two derivatives are continuous at the cutoff.

 All results are expressed in reduced Lennard–Jones units, with $\epsilon_{11}$ setting the energy scale, $\sigma_{11}$ the length scale, and $\sigma_{11}\sqrt{m/\epsilon_{11}}$ the time scale. The particle mass is set to $m=1$ for all species.

The system evolves in a square simulation box with $L_x = L_y$ and a constant particle density $\rho = N/(L_xL_y) = 1.19$, where $N$ denotes the total number of particles. We mainly consider a system with $N = 77$ particles, which corresponds to a box length of $L_x = 8.036$. We also study a larger system with $N = 704$ at relatively high temperatures, for which $L_x = 24.32$. The representative temperatures of this system are the onset temperature for glassy dynamics, $T_{\rm onset} \approx 0.5$, the mode-coupling transition temperature, $T_{\rm mct} \approx 0.3$, and the estimated empirical (experimental) glass transition temperature, $T_g \approx 0.15$~\cite{jung2025numerical}.

\subsection{Equilibration, inherent structures, and thermally-averaged configurations}

The equilibration procedure combines SWAP Monte Carlo (SMC)~\cite{swap:ninarello2017,swap:Berthier2019} with parallel tempering (PT) ~\cite{hukushima1996exchange} and population annealing (PA)~\cite{hukushima2003population}. The full procedure is described in detail in Ref.~\cite{jung2025numerical}, and we adopt here the same protocol and parameters as in that work. Importantly, we combine SMC with PT to create a set of roughly $N_s=2\cdot 10^5$ configurations at a temperature slightly below $T_g.$ These $N_s$ configurations are then used as input for a PA equilibration to reach very low temperatures. This procedure yields a set of $N_s$ equilibrated configurations at various temperatures, including temperatures down to $T \to 0$.

For each configuration $\alpha \in \{1,\dots,N_s\}$, characterized by the particle coordinates $\mathbf{r}_\alpha^N = (\mathbf{r}_{\alpha,1}, \mathbf{r}_{\alpha,2}, \dots, \mathbf{r}_{\alpha,N})$, we identify the corresponding nearest inherent structure, denoted by $\mathbf{r}_{\mathrm{inh},\alpha}^N$. This configuration is obtained by minimizing the potential energy
\[
E_{\rm pot}(\mathbf{r}^N) = \sum_{i<j} V(r_{ij}),
\]
using a steepest-descent algorithm with a maximal displacement per particle of $\Delta r = 0.01$ at each step. We then define the inherent-structure energy of configuration $\alpha$ as
\[
E_{\mathrm{inh},\alpha} = E_{\rm pot}(\mathbf{r}_{\mathrm{inh},\alpha}^N),
\]
which corresponds to the potential energy evaluated at the associated local minimum.

To calculate the thermally averaged configuration $\bar{\mathbf{r}}_\alpha^N$, we start from the configuration $\mathbf{r}_\alpha^N$ and perform $N_{\rm MC} = 7500$ Monte Carlo steps with a maximal displacement $\Delta r = 0.02$. This number of Monte Carlo steps is sufficient to determine thermally averaged local structures with reduced fluctuations, while avoiding large-scale rearrangements because the temperatures considered are very low, $T < T_g$. We then define
\[
\bar{\mathbf{r}}_\alpha^N = \frac{1}{N_{\rm MC}} \sum_{k=1}^{N_{\rm MC}} \mathbf{r}_\alpha^N(k).
\]

\subsection{Configurational Entropy and Rényi Complexity from Standard Methods}

We compute the configurational entropy $s^{\mathrm{conf}}$ and the $\mathrm{R\acute{e}nyi}$ complexity $s_m^{\mathrm{R\acute{e}nyi}}$ for $m=2,3,\ldots$ using conventional methods, i.e., without relying on the Grassberger-Procaccia construction.

\subsubsection{Inherent structures}

By grouping configurations according to their inherent-structure energies, as defined above, we construct a normalized probability distribution $\rho_\alpha^{\mathrm{inh}}$, where $\rho_\alpha^{\mathrm{inh}}$ denotes the fraction of independent configurations associated with the inherent-structure energy $E_{\mathrm{inh},\alpha}$. We then evaluate
\begin{align}
    s^{\mathrm{conf,inh}} &= - \frac{1}{N}\sum_\alpha \rho_\alpha^{\mathrm{inh}} \ln \rho_\alpha^{\mathrm{inh}}, \label{eq:sconf_inf}\\
    s_m^{\mathrm{R\acute{e}nyi,inh}} &= - \frac{1}{m-1} \frac{1}{N}\ln \sum_\alpha \left(\rho_\alpha^{\mathrm{inh}}\right)^m \label{eq:sm_inh},
\end{align}
where $m=2,3,...$.
As long as the set of inherent structures is sampled nearly completely, which is the case at very low temperatures, this definition provides a good measure of $s^{\mathrm{conf}}$ and $s_m^{\mathrm{R\acute{e}nyi}}$.

\subsection{Franz-Parisi potentials}

We also compute the Franz-Parisi potentials using both the quenched and annealed definitions. The quenched Franz-Parisi potential provides an alternative route to estimating the configurational entropy~\cite{berthier2019configurational}. The annealed Franz-Parisi potential, together with its generalization to $m$ clones, provides an independent estimate of the $\mathrm{R\acute{e}nyi}$ complexity of order $m$~\cite{javerzat2025renyi}.

\subsubsection{Quenched Franz-Parisi potential}

We described in detail in Ref.~\cite{jung2025numerical} how to extract the quenched Franz-Parisi potential,
\begin{equation}
\beta V^{\rm Quench}(Q) = - \frac{1}{N N_s}\sum_{\alpha=1}^{N_s}  \ln \left( \frac{1}{N_s} \sum_{\beta=1}^{N_s} \delta(Q-Q_{\alpha\beta}) \right).
\end{equation}
Here, the overlap $Q$ between two independently sampled configurations $\mathbf{r}_1^N$ and $\mathbf{r}_2^N$ is defined as
\begin{equation}
    Q(\mathcal{D}) = \frac{1}{N} \sum_{i=1}^N F\big( \mathcal{D}_{i} \big),
\end{equation}
where $\mathcal{D}=(\mathcal{D}_{1},\mathcal{D}_{2},\dots,\mathcal{D}_{N})$ denotes the set of minimal distances between the two configurations. These distances are determined by a brute-force procedure that explicitly accounts for all relevant symmetries; see Appendix VI of Ref.~\cite{jung2025numerical}. Here, we use the overlap function,
\begin{align}
    F_\text{mix}(x) &= \frac{1}{2} \left[  e^{ -\left(\frac{x}{a_0(T)} \right)^2 } +   e^{ -\left(\frac{x}{a_1}\right)^2 } \right],
\end{align}
with $a_1=0.1$ and the temperature-dependent length scale $a_0(T)$ is determined from the square root of the plateau value of the mean-squared displacement at temperature $T$.

Finally, we extract the configurational entropy using the relation
\begin{equation}\label{eq:sconf_fp}
s^{\text{conf,FP}} = \beta \left[ V^{\rm Quench}(Q_{\rm glass}) - V^{\rm Quench}(Q_{\rm liq}) \right],
\end{equation}
i.e., as the difference in the Franz-Parisi potential between the liquid minimum at low overlap, $Q_{\rm liq}$, and the glass minimum at high overlap, $Q_{\rm glass}$.

\subsubsection{Annealed Franz-Parisi potential}

Similarly, we define the generalized annealed Franz-Parisi potential~\cite{javerzat2025renyi} for $m$ clones as
\begin{equation}
\beta V^{\rm Anneal}_m(Q) = - \frac{1}{N} \ln \left( \frac{1}{N_s^m}\sum_{\alpha_1, \alpha_2,...,\alpha_m}  \prod_{a<b}^m \delta(Q-Q_{\alpha_a \alpha_b}) \right),
\end{equation}
using the same definition of the overlap $Q$ as above. From this quantity, we estimate the $\mathrm{R\acute{e}nyi}$ complexity via
\begin{equation}
\label{eq:sm_fp}
s_m^\mathrm{R\acute{e}nyi,FP} = \frac{1}{m-1}\beta \left[ V^{\rm Anneal}_m(Q_{\rm glass}) - V^{\rm Anneal}_m(Q_{\rm liq}) \right].
\end{equation}

\subsection{Aging data}

The thermal equilibration procedure based on population annealing depends sensitively on the number of initial configurations, $N_s$, used as input. If $N_s$ is too small, the ensemble eventually falls out of equilibrium.

To study the configurational entropy during aging in Fig.~\ref{fig:aging}, we construct ensembles in which $N_s$ is intentionally restricted, thereby generating nonequilibrium samples that mimic glasses at different stages of aging. Small $N_s$ corresponds to short aging times, whereas large $N_s$ corresponds to better-annealed glasses. To generate these nonequilibrium ensembles, we perform $N_{\mathrm{PA}}$ independent population-annealing runs starting from $N_s$ randomly chosen configurations. Typical values range from $N_{\mathrm{PA}} = 1024$ for $N_s = 2$ to $N_{\mathrm{PA}} = 32$ for $N_s = 2000$. We then combine the resulting $N_{\mathrm{PA}}N_s$ configurations and analyze their properties in the same way as for the fully equilibrated samples. This procedure is identical to that used in Fig.~1(d) of Ref.~\cite{jung2025numerical} to evaluate the ``time-dependent'' specific heat.

\section{Theoretical background of the Grassberger-Procaccia method}\label{app:derivation}

We explain why the Grassberger-Procaccia algorithm can be used to estimate the configurational entropy and $\mathrm{R\acute{e}nyi}$ complexity of glass-forming liquids within a statistical-mechanical framework.

\subsection{Configurational entropy}

We first recall the standard definition of the configurational entropy, also referred to as the \textit{complexity}, in order to introduce the notation used in this work. 
The configurational entropy $s^{\mathrm{conf}}$ is defined as
\begin{eqnarray}
S^{\mathrm{conf}} 
&=& S^{\mathrm{liq}} - S^{\mathrm{glass}} \nonumber \\
&=& \beta \left( E^{\mathrm{liq}} - F^{\mathrm{liq}} \right) 
     - \beta \left( E^{\mathrm{glass}} - F^{\mathrm{glass}} \right) \nonumber \\
&\simeq& -\beta \left( F^{\mathrm{liq}} - F^{\mathrm{glass}} \right),
\label{eq:S_conf_def}
\end{eqnarray}
where $S^{\mathrm{liq}}$ and $S^{\mathrm{glass}}$ denote the entropies of the liquid and glass states, respectively.
The last approximation assumes that the internal energies of the liquid and glass states, 
$E^{\mathrm{liq}}$ and $E^{\mathrm{glass}}$, are nearly identical. 
Consequently, the configurational entropy $S^{\mathrm{conf}}$ is primarily determined by the difference 
between the free energies of the liquid ($F^{\mathrm{liq}}$) and the glass ($F^{\mathrm{glass}}$).

For the liquid free energy, one obtains 
\[
\beta F^{\mathrm{liq}} = -\ln Z,
\]
where $Z$ is the total partition function of the system.
For the glass state, particularly at low temperatures, the partition function can be decomposed into contributions from individual metastable states labeled by $\alpha$:
\[
Z \simeq \sum_{\alpha} Z_{\alpha},
\]
where $Z_{\alpha}$ is the partition function associated with a metastable state $\alpha$. 
The quantity 
\[
\rho_{\alpha} = \frac{Z_{\alpha}}{Z}
\]
then represents the probability of finding the system in the metastable state $\alpha$.
Using these definitions, the free energy of the glass state can be expressed as
\begin{equation}
\beta F^{\mathrm{glass}} = -\sum_{\alpha} \rho_{\alpha} \ln Z_{\alpha}.
\end{equation}

Hence, $S^{\rm conf}$ in Eq.~(\ref{eq:S_conf_def}) becomes
\begin{eqnarray}
S^{\mathrm{conf}} 
&\simeq& \ln Z - \sum_{\alpha} \rho_{\alpha} \ln Z_{\alpha} \nonumber \\
&=& -\sum_{\alpha} \rho_{\alpha} \ln \rho_{\alpha}.
\label{eq:S_conf_Shannon}
\end{eqnarray}
Thus, $S^{\mathrm{conf}}$ can be interpreted as the Shannon entropy of the probability distribution 
$\rho_{\alpha}$.

\subsection{Rényi complexity}

The Rényi entropy is a one-parameter extension of the Shannon entropy~\cite{renyi1961measures} that has been widely used across many domains in science and engineering (see Ref.~\cite{ozawa2024perspective} for review). 
The configurational entropy $S^{\mathrm{conf}}$ in Eq.~(\ref{eq:S_conf_Shannon}) can be generalized to the Rényi form with an index $m \ge 0$, 
which we refer to as the \textit{Rényi complexity}:
\begin{equation}
    S^{\mathrm{R\acute{e}nyi}}_{m} 
    = - \frac{1}{m - 1} 
      \ln \sum_{\alpha} \left( \rho_{\alpha} \right)^{m}.
      \label{eq:Reny_def}
\end{equation}
It is straightforward to verify that the limit $m \to 1$ recovers the Shannon entropy, namely,
\begin{equation}
    \lim_{m \to 1} S^{\mathrm{R\acute{e}nyi}}_{m} 
    = S^{\mathrm{conf}} .
    \label{eq:Shannon_limit}
\end{equation}
Moreover, one can show that $S^{\mathrm{R\acute{e}nyi}}_{m}$ is a non-increasing function of $m$, 
and therefore, in general,
\begin{equation}
    S^{\mathrm{conf}} \ge S^{\mathrm{R\acute{e}nyi}}_{m} 
    \quad \text{for } m \ge 1.
\end{equation}

Furthermore, the Rényi complexity is closely related to the Franz–Parisi potential, 
which provides an alternative definition of the configurational entropy. 
In particular, $S^{\mathrm{R\acute{e}nyi}}_{m}$ can be expressed in terms of the $m$-component annealed Franz–Parisi potential (difference)
$V^{\mathrm{Anneal}}_{m}$ as~\cite{javerzat2025renyi}
\begin{equation}
    \frac{S^{\mathrm{R\acute{e}nyi}}_{m}}{N}
    = \frac{1}{m - 1} \, \beta V^{\mathrm{Anneal}}_{m}.
\end{equation}
It has been shown that, in mean-field spin glass models, 
$S^{\mathrm{R\acute{e}nyi}}_{m}/N$ for $m \ge 1$ vanishes at the same Kauzmann transition temperature $T_{\mathrm{K}}$, 
independently of $m$~\cite{javerzat2025renyi}. 
This observation suggests that $S^{\mathrm{R\acute{e}nyi}}_{m}$ serves as an alternative and useful quantity 
to locate or estimate the putative $T_{\mathrm{K}}$ and to study the nature of the Kauzmann transition.

\subsection{Measurement of Rényi complexity}

Compared with the Shannon form in Eq.~(\ref{eq:S_conf_Shannon}), 
the Rényi entropy $S^{\mathrm{R\acute{e}nyi}}_{m}$ in Eq.~(\ref{eq:Reny_def}) 
involves the functional form $\sum_{\alpha} (\rho_{\alpha})^{m}$.
This form is much easier to compute or estimate than the Shannon expression $\sum_\alpha \rho_{\alpha} \ln \rho_{\alpha}$.
Such accessibility makes the Rényi entropy particularly attractive and widely used in various domains of physics, 
including quantum many-body systems, where it is employed to quantify the magnitude of entanglement. 
In this work, our main focus is therefore to evaluate $\sum_{\alpha} (\rho_{\alpha})^{m}$ 
and to establish its connection with the Grassberger–Procaccia algorithm~\cite{grassberger1983estimation}.

We first relabel each metastable state by its thermally averaged particle positions, rather than by an abstract index $\alpha$.
In a metastable glass state, particles vibrate around their mean positions on the vibrational timescale $\tau_{\mathrm{vib}}$.
We therefore define the thermally averaged configuration
\[
\bar{\mathbf{r}}^{\,N} = (\bar{\mathbf{r}}_1, \bar{\mathbf{r}}_2, \ldots, \bar{\mathbf{r}}_N),
\qquad
\bar{\mathbf{r}}_i = \frac{1}{\tau_{\mathrm{vib}}} \int_0^{\tau_{\mathrm{vib}}} \mathbf{r}_i(t)\, dt.
\]
In this representation, each metastable state is specified by a unique set of averaged positions $\bar{\mathbf{r}}^{\,N}$, 
and the corresponding probability density is denoted by $\rho(\bar{\mathbf{r}}^{\,N})$. 
Our goal is therefore to estimate
\begin{equation}
    S^{\mathrm{R\acute{e}nyi}}_{m} 
    = - \frac{1}{m - 1}
      \ln \sum_{\bar{\mathbf{r}}^{\,N}} 
      \left( \rho(\bar{\mathbf{r}}^{\,N}) \right)^m ,
    \label{eq:SR_def}
\end{equation}
in particular, the term $\sum_{\bar{\mathbf{r}}^{\,N}} \left( \rho(\bar{\mathbf{r}}^{\,N}) \right)^m$.

Let us now reinterpret the meaning of this expression using the notion of $m$ replicas (or $m$ independent copies of the system), 
which provides a natural path to connect with the Grassberger--Procaccia algorithm~\cite{kurchan2010order}. 
The quantity $\rho(\bar{\mathbf{r}}^{\,N})$ represents the probability of finding a metastable state 
characterized by the thermally averaged configuration $\bar{\mathbf{r}}^{\,N}$ of size $N$. 
Then, $\left(\rho(\bar{\mathbf{r}}^{\,N})\right)^m$ can be interpreted as the probability that 
$m$ independent replicas of the system all occupy the same metastable state 
specified by $\bar{\mathbf{r}}^{\,N}$. 
Consequently, the sum $\sum_{\bar{\mathbf{r}}^{\,N}} \left( \rho(\bar{\mathbf{r}}^{\,N}) \right)^m$
corresponds to the total probability that all $m$ replicas are found in the \emph{same} metastable state 
(without specifying which one). 

Thus, the key quantity to evaluate is the probability that different replicas of the system 
share the same thermally averaged configuration $\bar{\mathbf{r}}^{\,N}$.
However, direct evaluation of this probability becomes computationally intractable when $N$ is large.

To make the problem numerically accessible, 
we consider a subset of $N$ particles forming a smaller patch of size $d \le N$, 
and study the limit $d \to N$. 
In this reduced formulation, we approximate
\begin{equation}
    \sum_{\bar{\mathbf{r}}^{\,N}} 
    \left( \rho(\bar{\mathbf{r}}^{\,N}) \right)^m
    = 
    \lim_{d \to N} 
    \sum_{\bar{\mathbf{r}}^{\,d}} 
    \left( \rho(\bar{\mathbf{r}}^{\,d}) \right)^m .
\end{equation}

Let us now consider the case of $m = 2$ in detail. 
As discussed above, $\left( \rho(\bar{\mathbf{r}}^{\,d}) \right)^2$ measures the probability that two replicas of size $d$
occupy the same thermally averaged configuration $\bar{\mathbf{r}}^{\,d}$. 
This viewpoint effectively compares two configurations of size $d$.

We now reformulate the problem as follows. 
Consider a large configuration consisting of $N \gg 1$ particles. 
Each particle $i$ (for $i = 1, 2, \ldots, N$) is taken as the center of a local patch that contains $d$ particles (including itself). 
Thus, each particle defines its own patch, and we can identify the configuration index $\bar{\mathbf{r}}^{\,d}$ with the particle index $i$.
We then scan over all other $N-1$ patches and count how many times a patch identical to that of particle $i$ is found. 
The fraction of such occurrences is denoted by
\[
f_i^{(d)} = \frac{n_i^{(d)}}{N - 1} \simeq \frac{n_i^{(d)}}{N},
\]
where $n_i^{(d)}$ is the number of patches that are identical to the patch centered at particle $i$.

In practice, $n_i^{(d)}$ can be computed as
\begin{equation}\label{eq:ni}
    n_i^{(d)}(\epsilon) = \sum_{\substack{j=1 \\ (j\neq i)}}^{N} \theta\!\left( \epsilon - \Delta_{ij}^{(d)} \right),
\end{equation}
where $\theta(\cdot)$ is the Heaviside step function, and 
$\Delta_{ij}^{(d)}$ denotes a suitably defined ``distance'' between patches $i$ and $j$ 
(the precise definition is provided in Appendix~\ref{app:shape_matching}). 
The parameter $\epsilon$ serves as a resolution threshold that determines whether two patches 
are considered identical within tolerance $\epsilon$. 
Hence, $f_i^{(d)}$ (and thus $n_i^{(d)}$) is a directly measurable quantity, provided that 
$\Delta_{ij}^{(d)}$ is properly defined.

Using this patch representation, the quantity $\sum_{\bar{\mathbf{r}}^{\,d}} (\rho(\bar{\mathbf{r}}^{\,d}))^2$, which is the average of $\rho(\bar{\mathbf{r}}^{\,d})$,  
can be estimated as the average of $f_i^{(d)}$ for a given resolution $\epsilon$:
\begin{eqnarray}
    \sum_{\bar{\mathbf{r}}^{\,d}} 
    \left( \rho(\bar{\mathbf{r}}^{\,d}) \right)^2
    &=& 
    \sum_{\bar{\mathbf{r}}^{\,d}} 
    \rho(\bar{\mathbf{r}}^{\,d}) \cdot \rho(\bar{\mathbf{r}}^{\,d})
    \nonumber \\
    &\simeq& 
    \frac{1}{N} \sum_{i=1}^{N} f_i^{(d)} 
    \simeq 
    \frac{1}{N^2} \sum_{i=1}^{N} n_i^{(d)}
    \nonumber \\
    &=& 
    \frac{1}{N^2} \sum_{i=1}^{N} \sum_{\substack{j=1 \\ (j\neq i)}}^{N} 
    \theta\!\left( \epsilon - \Delta_{ij}^{(d)} \right)
    \nonumber \\
    &=& 
    \frac{
    \left[ \text{Number of pairs $(i,j)$ with } 
    \Delta_{ij}^{(d)} < \epsilon \right]
    }{N^2}. \nonumber \\
    \label{eq:m2_estimator}
\end{eqnarray}
Equation~(\ref{eq:m2_estimator}) thus provides a practical estimator 
for $\sum_{\bar{\mathbf{r}}^{\,d}} (\rho(\bar{\mathbf{r}}^{\,d}))^m$ in the case $m = 2$.

For more general values of $m$, 
we can similarly evaluate $\sum_{\bar{\mathbf{r}}^{\,d}} (\rho(\bar{\mathbf{r}}^{\,d}))^m$ 
in terms of the average of $(\rho(\bar{\mathbf{r}}^{\,d}))^{m-1}$, namely,
\begin{eqnarray}
    \sum_{\bar{\mathbf{r}}^{\,d}} 
    \left( \rho(\bar{\mathbf{r}}^{\,d}) \right)^{m}
    &=& 
    \sum_{\bar{\mathbf{r}}^{\,d}} 
    \rho(\bar{\mathbf{r}}^{\,d})
    \left( \rho(\bar{\mathbf{r}}^{\,d}) \right)^{m-1}
    \nonumber \\
    &\simeq& 
    \frac{1}{N} \sum_{i=1}^{N} 
    \left( f_i^{(d)}(\epsilon) \right)^{m-1}
    \nonumber \\
    &\simeq& 
    \frac{1}{N^{m}} \sum_{i=1}^{N} 
    \left( n_i^{(d)}(\epsilon) \right)^{m-1} \nonumber \\
    &=& \frac{1}{N^{m-1}} C_m^{(d)}(\epsilon) ,
    \label{eq:m_general_estimator}
\end{eqnarray}
where $C_m^{(d)}(\epsilon)$ is a correlation integral defined as
\begin{equation}
    C_m^{(d)}(\epsilon) = \frac{1}{N} \sum_{i=1}^N \left[ n_i^{(d)}(\epsilon)\right]^{m-1} .
    \label{eq:correlation_integral_SI}
\end{equation}

Equation~(\ref{eq:m_general_estimator}) shows that 
$\sum_{\bar{\mathbf{r}}^{\,d}} (\rho(\bar{\mathbf{r}}^{\,d}))^m$ 
can be numerically estimated once the patch similarity counts $n_i^{(d)}$ are computed. 

In the limit $\epsilon \to 0$ and for $N \gg 1$, the intensive $\mathrm{R\acute{e}nyi}$ complexity $s^{\mathrm{R\acute{e}nyi}}_{m}$ is computed as\begin{eqnarray}
s^{\mathrm{R\acute{e}nyi}}_{m} 
&=& \frac{S^{\mathrm{R\acute{e}nyi}}_{m}}{N} 
= \lim_{d \to N} 
   \frac{1}{d(1 - m)} 
   \ln \sum_{\bar{\mathbf{r}}^{\,d}} 
   \left( \rho(\bar{\mathbf{r}}^{\,d}) \right)^{m}
   \nonumber \\
&\simeq& \lim_{\epsilon\to0}\,
\lim_{d \to N} 
\frac{1}{d(1 - m)} 
\ln C_m^{(d)}(\epsilon) .
\label{eq:s_reny_intensive}
\end{eqnarray}
Thus, $s^{\mathrm{R\acute{e}nyi}}_{m}$ can be extracted from the asymptotic behavior for sufficiently small $\epsilon$ and sufficiently large $d$.

\subsection{The $m \to 1$ limit: Configurational entropy}

In the limit $m \to 1$, the R\'enyi complexity reduces to the configurational entropy, as shown in Eq.~(\ref{eq:Shannon_limit}). We now derive the corresponding expression for the configurational entropy using the correlation integral~\cite{cohen1985computing}.

First, for small $m-1$, the correlation integral $C_m^{(d)}(\epsilon)$ in Eq.~(\ref{eq:correlation_integral_SI}) can be expanded as
\begin{eqnarray}
    C_m^{(d)}(\epsilon) &=& \frac{1}{N} \sum_{i=1}^N \exp\!\left[(m-1)\ln n_i^{(d)}(\epsilon)\right] \nonumber \\
    &=& 1 + \frac{m-1}{N}\sum_{i=1}^N \ln n_i^{(d)}(\epsilon) + \mathcal{O}\left((m-1)^2\right) . \nonumber
\end{eqnarray}
We then obtain
\begin{equation}
    \ln C_m^{(d)}(\epsilon) = \frac{m-1}{N}\sum_{i=1}^N \ln n_i^{(d)}(\epsilon) + \mathcal{O}\left((m-1)^2\right) .
\end{equation}

Using Eq.~(\ref{eq:s_reny_intensive}), the configurational entropy, corresponding to the limit $m \to 1$, can be obtained as
\begin{eqnarray}
    s^{\rm conf} &=& \lim_{m \to 1} s_m^{\mathrm{R\acute{e}nyi}} \nonumber \\
    &\simeq& \lim_{\epsilon \to 0} \lim_{d \to \infty} \lim_{m \to 1}
    \frac{1}{d(1-m)} \ln C_m^{(d)}(\epsilon) \nonumber \\
    &=& - \lim_{\epsilon \to 0} \lim_{d \to \infty} \frac{1}{d}
    \ln \tilde C_1^{(d)}(\epsilon),
\end{eqnarray}
where we define
\begin{equation}
    \ln \tilde C_1^{(d)}(\epsilon)
    = \frac{1}{N} \sum_{i=1}^N \ln n_i^{(d)}(\epsilon) .
\end{equation}
Therefore, $s^{\rm conf}$ can be extracted from the asymptotic behavior of $\ln \tilde C_1^{(d)}(\epsilon)$ in the limits $\epsilon \to 0$ and $d \to \infty$.

Note that for a sufficiently large system, one expects $n_i^{(d)}(\epsilon)\geq 1$. However, to estimate the configurational entropy for our $N=77$ system we discard the terms $n_i^{(d)}=0$ in the above sum. Alternatively, one may redefine $n_i^{(d)}$ so that each patch is counted as matching itself, namely by including the $i=j$ terms in the summation in Eq.~(\ref{eq:ni1}). We have checked that the resulting correlation integrals behaves similarly.

\section{Patch matching algorithm}\label{app:shape_matching}

Identifying similar configurational patches in amorphous structures is a nontrivial task due to the intrinsically disordered nature of the configurations. 
In essence, the problem reduces to computing the distance (e.g., root-mean-square deviation, RMSD) between two patches, $A$ and $B$, each containing $d$ particles. 
Because the patches are embedded in a homogeneous but disordered environment, 
it is necessary to determine the optimal rotation and translation that best align $A$ and $B$ to evaluate their distance. 
The Kabsch algorithm~\cite{Kabsch1976} (or Procrustes analysis), widely used in computer vision and structural biology, provides a means to obtain the optimal rotation and translation by means of a singular value decomposition (SVD). 
However, an important prerequisite for applying the Kabsch algorithm is that the correspondence between particles in patches $A$ and $B$ is known -
that is, one must determine which particle in $A$ should be compared with which particle in $B$. This constitutes a pair-assignment problem, which is particularly challenging in amorphous configurations.

Previous studies on patch identification in amorphous systems typically optimized the rotational alignment in a discretized manner~\cite{sausset2011characterizing},  or alternatively employed the Kabsch algorithm to obtain optimal rotation and translation in a continuous manner but relied on heuristic approaches to resolve the pair-assignment problem~\cite{hallett2018local}. 
In contrast, the algorithm used in this work improves the overall accuracy by 
simultaneously addressing both rotation–translation optimization and particle assignment in a systematic way. The method is based on a recently developed algorithm~\cite{Gunde2021,these_miha} originally designed for shape matching between distorted atomic structures, that we have adapted to the case of amorphous structures.
This higher precision makes our method particularly suitable for the evaluation of the Rényi entropy, which requires accurate estimation of configurational similarities.
The source codes implementing our algorithm are openly available at \cite{Gitcode}.

\subsection{Notations}

We first define a patch of size $d$ centered on a reference particle $i$ as the set of particle positions
\[
A_{i}^{(d)} = \left\{ \mathbf{r}_{1}^A = \mathbf{r}_{i}^A, \mathbf{r}_{2}^A, \ldots, \mathbf{r}_{d}^A \right\},
\]
where the $d$ particles are chosen as those located closest to the reference particle $i$. We denote $\tilde{\mathbf{r}}_{k}^A$ the coordinates relative to the reference (central) particle of the patch: $\tilde{\mathbf{r}}^A_{k} = \mathbf{r}^A_{k} - \mathbf{r}_1^A$.
For two patches of equal size $d$, denoted $A_{i}^{(d)}$ and $B_{j}^{(d)}$, 
an \emph{assignment} $\sigma$ is defined as a one-to-one correspondence 
that maps each particle index in patch $A$ to a particle index in patch $B$, namely,
\begin{align*}
    \sigma : &\; A \rightarrow B, \\
    &\; k \mapsto \sigma(k).
\end{align*}
This mapping specifies which particle in patch $A$ is to be compared with which particle in patch $B$ 
when computing their relative distance.
Given $A_{i}^{(d)}$, $B_{j}^{(d)}$, an assignment $\sigma$, and a rotation matrix $R$, we define the mutual distance $\Delta(A,B; \sigma,R)$ as the root mean-square deviation (up to a factor $\sqrt{d}$)
\begin{equation}\label{eq:deltadef}
    \Delta(A_{i}^{(d)},B_{j}^{(d)};\sigma,R) \equiv \sqrt{\sum_{k=1}^d \left|\tilde{\mathbf{r}}_{k}^A -R \,\tilde{\mathbf{r}}_{\sigma(k)}^B\right|^2} .
\end{equation}
Such distance depends on the chosen assignment and rotation matrix. We then define the patch distance $\Delta_{i,j}^{(d)}$ entering Eq.~(\ref{eq:ni}) as the minimal such distance, namely
\begin{equation}\label{eq:minpb}
    \Delta_{i,j}^{(d)} = \min_{\left\{\sigma, R\right\}}\Delta(A_i^{(d)},B_j^{(d)};\sigma, R) .
\end{equation}
We denote $\sigma_{\rm opt}$, $R_{\rm opt}$ the assignment and rotation matrix that are solution to Eq.~(\ref{eq:minpb}). 
We note that the optimization with respect to translation is not written explicitly, 
as it is handled within the Kabsch algorithm.

Our patch matching algorithm, whose aim is to determine $\sigma_{\rm opt}$, $R_{\rm opt}$ and in fine $\Delta$, is based on the Iterative Rotations and Assignments (IRA) method \cite{Gunde2021,these_miha}, that has been designed to solve the double optimization problem in Eq. (\ref{eq:minpb}) for pairs of slightly distorted atomic structures. We first give a brief review of IRA, before adapting it to amorphous systems. In this work we restrict to 2D, but the method can be generalized to higher spatial dimensions.

\subsection{Shape matching by IRA for atomic structures}
To simultaneously optimize rotation and assignment, the algorithm~\cite{Gunde2021} goes in two stages: first roughly scan the rotation space and select the assignment that minimizes the distance between rotated patches; second, fix the assignment and refine the rotation using the Kabsch algorithm.

Given two atomic structures $A$ and $B$, 
the rotational space is first discretized using a basis defined by atomic vectors. 
For each basis element $\alpha$, corresponding to a discrete rotation $R_{\alpha}$, 
the optimal particle assignment $\sigma$ is determined by minimizing the Hausdorff distance 
$h(A, R_{\alpha} B)$ between structure $A$ and the rotated structure $R_{\alpha} B$. 
The (permutation-invariant) Hausdorff distance is defined as
\begin{equation}
h(A,B) = \max_i \, \min_j \left| \mathbf{r}_i^A - \mathbf{r}_j^B \right|,
\end{equation}
and is computed from the all-to-all distance matrix 
$d_{ij} = \left| \mathbf{r}_i^A - \mathbf{r}_j^B \right|$ 
using the \emph{Constrained Shortest Distance Assignment} (CShDA) algorithm 
(see Ref.~\cite{these_miha}). 
The basis element $\alpha_{\mathrm{min}}$ that yields the smallest Hausdorff distance, 
$h_{\mathrm{min}} = h(A, R_{\alpha_{\mathrm{min}}} B)$, 
defines the approximately optimal rotation 
$R_{\mathrm{approx}} = R_{\alpha_{\mathrm{min}}}$ 
and the corresponding optimal assignment $\sigma_{\mathrm{opt}}$.

Given the optimal assignment $\sigma_{\mathrm{opt}}$, 
the optimal rotation matrix $R_{\mathrm{opt}}$ is defined as the one that minimizes 
the RMSD between the sets of points $A$ and 
$\sigma_{\mathrm{opt}}\!\left( R_{\mathrm{approx}} B \right)$. 
The rotation $R_{\mathrm{opt}}$ is obtained using the standard Kabsch algorithm~\cite{Kabsch1976}, 
which determines the optimal rotation and translation in a continuous manner 
via singular value decomposition (SVD), without resorting to any discretization.
The distance between patches A and B is then taken as
\begin{equation}
    \Delta_{\rm IRA}\left(A_i^{(d)},B_j^{(d)}\right) = \sqrt{\sum_{k=1}^d \left|\tilde{\mathbf{r}}_{i_k}^A -R_{\rm opt}\,R_{\rm approx} \,\tilde{\mathbf{r}}_{j_{\sigma_{\rm opt}(k)}}^B\right|^2}
\end{equation}
and it is conjectured that $\Delta_{\rm IRA}(A_i^{(d)},B_j^{(d)}) =  \displaystyle\min_{\left\{\sigma, R\right\}}\Delta(A_i^{(d)},B_j^{(d)};\sigma, R)$~\cite{Gunde2021}.

\subsection{Shape matching for amorphous patches}

The original Iterative Rotations and Assignments (IRA) method~\cite{Gunde2021} was designed primarily for comparing atomic structures that are fairly similar, for instance configurations related by small perturbations. In such systems, a good match between only a few particle pairs is typically sufficient 
to ensure structural similarity between configurations $A$ and $B$, 
so that a coarse discretization of the rotational space over a few atomic bases is adequate. 
This assumption, however, no longer holds for amorphous patches $A$ and $B$. 
In this case, the discretization of the rotational space must be performed on a finer grid 
to properly capture the structural variability. 
We therefore adapt the IRA procedure accordingly, as summarized in Algorithm~\ref{alg:IRA}. 
Specifically, patch $B$ is first transformed using a discrete set of improper rotations $R_k^{\pm}$ built from $M=250$ uniformly spaced orientations, starting from angle $\phi=0$ and incrementing the angle by $\delta\phi = 2\pi/M$. The set includes both rotations and their corresponding reflections.

In summary, the overall algorithm first determines the optimal particle-to-particle assignment 
under a discretized (but sufficiently fine) set of rotations, 
and subsequently applies the Kabsch algorithm to ``fine-tune'' the alignment, 
thereby obtaining the truly optimal rotation and translation.

We note that, in Appendix~\ref{app:derivation}, the general formulation is presented for patch identification within a single large system of $N$ particles. In our practical implementation, however, we consider $N_s$ independent configurations of finite size $N$ and identify patches among the resulting total of $N\times N_s$ particles.
We also note that particles in patch $A$ may sometimes best match particles that lie slightly outside 
the region containing the $d$ particles of patch $B$. 
To account for this possibility, we use a larger patch size for $B$, denoted by
$d' = d + \Delta d$ (see Algorithm~\ref{alg:full}). The additional margin
$\Delta d$ expands the search space and increases the likelihood of identifying
the optimal assignment between the two patches. In the range of patch sizes
studied here, we set $\Delta d = 40$; for substantially larger patches, this
value should be adjusted accordingly.

It is also worth noting that our implementation does not distinguish between particles of different species, 
although the system under consideration is a three-component mixture. 
This choice is motivated by our goal to estimate the entropy solely from the spatial configuration 
(i.e., particle positions) without incorporating additional information about particle species. 
Moreover, it is well established that in dense amorphous systems, 
the positional and size (or species) degrees of freedom are strongly coupled~\cite{ninarello2017models}; 
therefore, incorporating positional information alone already indirectly accounts 
for compositional effects to a large extent.

\begin{algorithm}[H]
\caption{IRA for amorphous patches}\label{alg:IRA}
\KwIn{$A_i^{(d)}$, $B_j^{(d')}$ in their reference frames\;
$M=250$}
\KwOut{%$R_{\rm opt} =R_{\rm SVD, opt}\cdot R_{k,\rm opt}$; $\sigma_{\rm opt}$; 
$\Delta^{(d)}_{i,j} = \displaystyle\min_{\sigma, R} \Delta^{(d)}_{i,j}(\sigma,R)$}
$\Delta_{\rm min} = 100$\;
\For{$\phi_k = 0,2\pi/M,...,(M-1)2\pi/M$}{
    $R_k^{\pm} = \begin{pmatrix}
\cos\phi_k & -\sin\phi_k\\ \pm \,\sin\phi_k &  \pm \cos\phi_k\end{pmatrix}$\;
%(rotation by $\phi_k$ and reflections)\\
    $\sigma^k_{\rm opt} = {\texttt{CShDA}}\left[A^{(d)}_i, R_k\,B^{(d')}_j\right]$\;
    %(assignment defined by the Hausdorff distance $h(A^{(d)}_i, R_k\,B^{(d')}_j)$)\\
    
    $\Delta_k = \displaystyle\min_{R_{\rm SVD}} \Delta\left(A^{(d)}_i, \sigma^k_{\rm opt}\left(R_{\rm SVD}\,R_k\,B^{(d')}_j\right)\right) = \texttt{Kabsch}\left[A^{(d)}_i, \sigma^k_{\rm opt}\left(R_{\rm SVD}\,R_k\,B^{(d')}_j\right)\right]$\;
    \If{$\Delta_k < \Delta_{\rm min}$}{$\Delta_{\rm min} = \Delta_k$}
}
\Return $\Delta_{i,j}^{(d)} = \Delta_{\rm min}$\;
\end{algorithm}
The CShDA algorithm is described in details in Ref.~\cite{these_miha}, and our implementation \cite{Gitcode} follows closely the one published at \cite{miha_code}.

Importantly, we have validated Algorithm~\ref{alg:IRA} against exact solutions 
for small patch sizes ($d \le 10$), 
where the exact optimal assignment can be obtained by explicitly enumerating all possible permutations 
and applying the Kabsch algorithm to compute the corresponding RMSD for each case. 
Our algorithm successfully recovered the exact solution in all tested pairs that have $\Delta<1$, 
thereby validating the accuracy and robustness of the proposed method. Note that for some occurrences of highly dissimilar structures ($\Delta\gtrsim1$), our method overestimates the true RMSD. These errors are expected to have no impact on our results as we are interested in the scaling regime $\epsilon \ll 1$.

A possible direction for future work is to accelerate the computation by
pre-screening patch pairs, e.g., using coarse-grained structural descriptors, thereby
discarding pairs that are clearly non-matching before computing the full
distance $\Delta_{ij}^{(d)}$. This would allow the expensive matching procedure
to focus on the similar pairs that contribute to the relevant
small-$\epsilon$ scaling regime.

\subsection{Full algorithm to compute the matrix $\Delta^{(d)}$}

\begin{algorithm}[H]
\caption{Full algorithm}\label{alg:full}
\KwIn{$N_s$ configurations of size $N$\;
patch size $d$\;
patch size $d'=d+\Delta d$}
\KwOut{distance matrix $\Delta^{(d)}$\;}

\For{$I = 1,\cdots,N\times N_s$}{
\For{$J=1,\cdots,N\times N_s$}{
\If{$I\neq J$}{
$\Delta_{I,J}^{(d)} = \texttt{IRA}\left[A_I^{(d)},B_J^{(d')}\right]$

}
}
   }
\Return $\Delta^{(d)}$\;
\end{algorithm}

\section{Numerical determination of Rényi complexities and configurational entropy}
\label{app:numerical}

% In this section we give more details about our measurements, and about the fitting procedure used in section~\ref{sec:extrac_K}. 

% \begin{table*}[]
%     \centering
%     \begin{tabular}{c|c|c|c}
%        N  & C   &   T   &   d \\ \hline
%        77  &500 &   $\left\{0.025, 0.029, 0.035, 0.047, 0.052, 0.061, 0.069, 0.079, 0.084, 0.087\right\}$ &$\left\{13, 14,\cdots, 21\right\}$\\
%        704  &   50  &   ?   &   ?
%     \end{tabular}
%     \caption{Parameters used for the measurement of the distance matrix $\Delta^{(d)}$: number $N$ of particles per configuration, number of configurations $C$, values of the temperature $T$, number $d$ of particles in a patch. The number of configurations is chosen such that $\mathcal{N} = N\times C \gtrsim 35.10^3$.}
%     \label{tab:measurements}
% \end{table*}

\subsection{Correlation integrals}

\begin{figure*}[!ht]
    \centering
    \begin{subfigure}{.32\linewidth}
    \includegraphics[width=\linewidth]{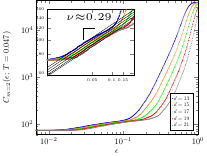}
    \end{subfigure}
    \hfill
    \begin{subfigure}{.32\linewidth}
    \includegraphics[width=\linewidth]{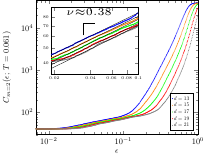}
    \end{subfigure}
    \hfill
    \begin{subfigure}{.32\linewidth}
    \includegraphics[width=\linewidth]{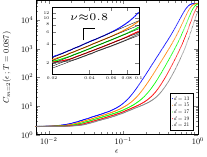}
\end{subfigure}
    \begin{subfigure}{.32\linewidth}
    \includegraphics[width=\linewidth]{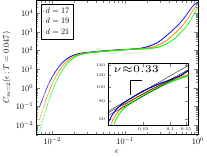}
    \end{subfigure}
    \hfill
    \begin{subfigure}{.32\linewidth}
    \includegraphics[width=\linewidth]{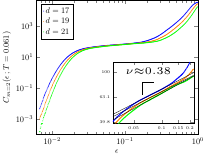}
    \end{subfigure}
    \hfill
    \begin{subfigure}{.32\linewidth}
    \includegraphics[width=\linewidth]{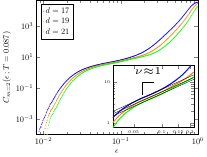}
    \end{subfigure}
    \caption{Correlation integral $C_m^{(d)}(\epsilon)$ for $m=2$ and several values of $d$, at temperatures $T=0.047$ (left), $T=0.061$ (center), and $T=0.087$ (right), computed from inherent structures (top row) and thermally averaged configurations (bottom row). The insets show a zoom of the scaling regime, where the correlation integral follows Eq.~(\ref{eq:GP_scaling}). The lines are fits to the form $\ln C_m = \nu \ln \epsilon + \mathrm{cst}$. }\label{fig:Cm}
\end{figure*}

Having identified the patches and computed the distance between pairs of patches, we evaluate the correlation integral $C_m^{(d)}(\epsilon)$ defined in Eq.~(\ref{eq:Cm1}). In Fig.~\ref{fig:Cm}, we show $C_m^{(d)}(\epsilon)$ for $m=2$, computed from two types of data: inherent structures and thermally averaged configurations.
We find that, in the small-$\epsilon$ regime, the correlation integral exhibits the scaling behavior expected from the Grassberger--Procaccia algorithm.

Note that, for the inherent-structure data, $C_m^{(d)}$ exhibits a plateau in the limit $\epsilon\to0$. This is because, at such low temperatures and for finite system size, several configurations may belong to the same inherent structure. The value of the plateau therefore depends on how many configurations fall into each inherent structure. Suppose that, among the $N_s$ configurations, $N_\alpha$ of them belong to the inherent structure labeled by $\alpha=1,2,\cdots,K$. Consider a patch centered at particle $I=(i,c)$, where configuration $c$ belongs to inherent structure $\alpha$. In the limit $\epsilon\to0$, the only patches $J=(j,c')$ that remain similar to patch $I$ are the $N_\alpha-1$ patches centered at $j=i$ in the other configurations belonging to the same inherent structure. Namely,
\begin{equation}
\begin{aligned}
    n_I^{(d)}(\epsilon\to0)
    &= \sum_{J=(j,c')} \Theta\!\left(-\Delta_{IJ}^{(d)}\right) \\
    &= \sum_{\alpha'=1}^K \sum_{\substack{c'\in {\rm IS}_{\alpha'}\\ c'\neq c}} \sum_{j\in c'}
    \delta_{\alpha',\alpha}\,\delta_{i,j} \\
    &= \sum_{\substack{c'\in {\rm IS}_{\alpha}\\ c'\neq c}} 1
    = N_\alpha-1.
\end{aligned}
\end{equation}
Therefore,
\begin{equation}
\begin{aligned}
    C_m^{(d)}(\epsilon\to0)
    &= \frac{1}{N\,N_s}\sum_{\alpha=1}^K \sum_{c\in {\rm IS}_{\alpha}} \sum_{i=1}^N
    \left[n_i^{(d)}(\epsilon\to0)\right]^{m-1} \\
    &= \frac{1}{N_s}\sum_{\alpha=1}^K N_\alpha (N_\alpha-1)^{m-1}.
\end{aligned}
\end{equation}
We checked the validity of the above result in the limit of very low temperature, $T\sim 0.001$, where all configurations fall into the same ground state. In this case, $K=1$ and $N_{\rm GS}=N_s$, so the plateau value becomes
\begin{equation}
    C_m^{(d)}(\epsilon\to0 \mid T\to0) = (N_s-1)^{m-1}.
\end{equation}
This prediction agrees with our data at $T=0.001$.

We also note that the plateau is absent in the correlation integrals computed from thermally averaged data, since in that case the configurations are close to, but always different from their inherent structure, even at very low temperature.

\begin{figure*}[!ht]
    \centering
    \begin{subfigure}{.32\linewidth}
    \includegraphics[width=\linewidth]{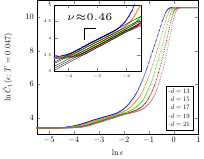}
    \end{subfigure}
    \hfill
    \begin{subfigure}{.32\linewidth}
    \includegraphics[width=\linewidth]{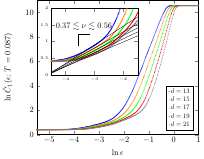}
    \end{subfigure}
    \hfill
    \begin{subfigure}{.32\linewidth}
    \includegraphics[width=\linewidth]{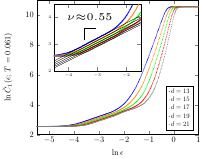}
\end{subfigure}
\caption{Correlation integral $\ln \tilde{C}_1$ [cf. Eq.~(\ref{eq:Ctilde})] as a function of $\ln \epsilon$ for $T=0.047$ (left), $T=0.061$ (center), and $T=0.087$ (right). The insets show a zoom of the scaling regime, where the black lines are fits to the form $\ln \tilde{C}_1 = \nu \ln \epsilon + b$. Note that the width of the scaling regime decreases for $T \gtrsim 0.08$.}\label{fig:C1}
\end{figure*}

\begin{figure*}[!ht]
    \centering
    \begin{subfigure}{.32\linewidth}
    \includegraphics[width=\linewidth]{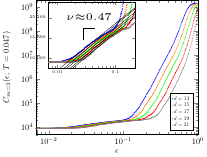}
    \end{subfigure}
    \hfill
    \begin{subfigure}{.32\linewidth}
    \includegraphics[width=\linewidth]{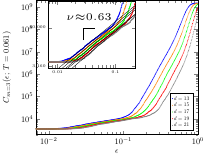}
    \end{subfigure}
    \hfill
    \begin{subfigure}{.32\linewidth}
    \includegraphics[width=\linewidth]{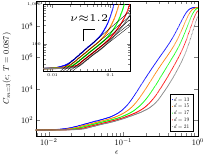}
\end{subfigure}
\caption{
Correlation integral $C_m^{(d)}(\epsilon)$ for $m=3$ and $d=13, 15, 17, 19, 21$, at temperatures $T=0.047$ (left), $T=0.061$ (center), and $T=0.087$ (right), computed from inherent structures. The insets show a zoom of the scaling regime, where the correlation integral follows Eq.~(\ref{eq:GP_scaling}). The lines are fits to the form $\ln C_m = \nu \ln \epsilon + \mathrm{cst}$.}\label{fig:C3}
\end{figure*}

In addition to the correlation integral for $m=2$, we also analyze the $m\to1$ case through the correlation integral $\tilde{C}_1$ shown in Fig.~\ref{fig:C1}, as well as the $m=3$ case shown in Fig.~\ref{fig:C3}. These data display the same qualitative behavior as for $m=2$, with the expected scaling behavior emerging in the small-$\epsilon$ regime.

\subsection{Finite size effects}

We have presented results for the system size $N=77$, for which very low-temperature equilibrium configurations are available. Here, we briefly discuss finite-size effects.

\begin{figure}
    \centering
    \includegraphics[width=\linewidth]{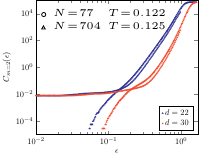}
    \caption{Correlation integral for systems of size $N=77$ and $N=704$ at comparable
temperatures and for patch sizes $d=22$ and $d=30$. The comparison illustrates
finite-size effects.}
    \label{fig:finite_size_effects}
\end{figure}

Figure~\ref{fig:finite_size_effects} shows $C_m^{(d)}(\epsilon)$ for $m=2$ for both $N=77$ and $N=704$ at similar temperatures, $T \approx 0.12$, and for two patch sizes, $d=22$ and $d=30$. Note that this temperature is the lowest for which equilibrated samples are obtained at $N=704$, but is relatively high compared with those studied in the main text.

At large $\epsilon$, the two system sizes show consistent behavior. At smaller $\epsilon$, however, the $N=77$ data exhibit a plateau as $\epsilon \to 0$, reflecting the fact that some independently sampled configurations belong to the same inherent structure. By contrast, the $N=704$ data continue to decrease upon lowering $\epsilon$, because in such a larger system no two sampled configurations belong to the same inherent structure. Importantly, in this temperature range, no clear scaling regime is observed for
either system size. Entropy estimation from patch statistics is therefore
primarily limited by the temperature range over which samples can be equilibrated.

\subsection{Thermal effects}

\begin{figure}
\includegraphics[width=\linewidth]{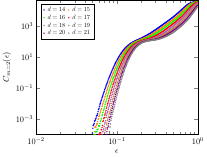}
 \includegraphics[width=\linewidth]{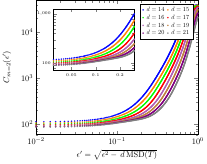}
 %\includegraphics[width=\linewidth]{figures/bare_replot.pdf}
    % \centering
    % \begin{subfigure}{.49\linewidth}
    %     \includegraphics[width=\linewidth]{figures/bare.pdf}
    % \end{subfigure}
    % \hfill
    % \begin{subfigure}{.49\linewidth}
    %     \includegraphics[width=\linewidth]{figures/bare_replot.pdf}
    % \end{subfigure}
    \caption{Correlation integral for instantaneous equilibrium configurations.
(a) $C_m^{(d)}$ as a function of the threshold $\epsilon$ at temperature
$T=0.047$. 
(b) The same quantity plotted as a function of the effective threshold
$\epsilon'$, where the cage size is measured from the plateau of the
mean-squared displacement (MSD) in independent Monte Carlo simulations. The inset
shows a zoom of the expected scaling regime, illustrating that an approximately linear
behavior emerges as $d$ increases.}
    \label{fig:thermal_effects}
\end{figure}

We discuss the role of short-time thermal averaging and inherent structures by
comparing them with bare instantaneous equilibrium configurations.
Figure~\ref{fig:thermal_effects}(a) shows the correlation integral for $m=2$
computed from instantaneous equilibrium configurations. In contrast to the
thermally averaged and inherent-structure configurations, the correlation
integral for instantaneous configurations decays rapidly to zero and does not
exhibit a clear scaling regime suitable for entropy extraction. This
demonstrates that removing thermal fluctuations, either through
inherent-structure analysis or short-time thermal averaging, is essential for
revealing the underlying structural backbone relevant to the entropy.

We next examine the role of thermal fluctuations in more detail.
We denote by $\Delta_{IJ}^{(d)}$ the RMSD between two patches $I$ and $J$ extracted from instantaneous thermal equilibrium configurations:
\begin{equation}
    \Delta_{IJ}^{(d)} = \sqrt{\sum_{k=1}^d \left| {\bf r}_k^I-{\bf r}_k^J \right|^2},
\end{equation}
where ${\bf r}_k^I$ and ${\bf r}_k^J$ are the positions of the $k$-th particles in patches $I$ and $J$, respectively, measured relative to their reference particles in the instantaneous configurations. We assume that the proper particle assignment and rotation have already been performed.

We decompose these positions into time-averaged positions and thermal fluctuations as
\begin{eqnarray}
    {\bf r}_k^I &=& \bar{\bf r}_k^I + \delta {\bf r}_k^I, \\
    {\bf r}_k^J &=& \bar{\bf r}_k^J + \delta {\bf r}_k^J.
\end{eqnarray}
With this decomposition, we obtain
\begin{eqnarray}
    \left[\Delta_{IJ}^{(d)}\right]^2 &=& \left[\bar{\Delta}_{IJ}^{(d)}\right]^2 
    + 2\sum_{k=1}^d (\bar{\bf r}_k^I-\bar{\bf r}_k^J)\cdot(\delta{\bf r}_k^I-\delta{\bf r}_k^J) \nonumber \\
    && + \sum_{k=1}^d |\delta{\bf r}_k^I|^2
    - 2\sum_{k=1}^d \delta{\bf r}_k^I \cdot \delta{\bf r}_k^J
    + \sum_{k=1}^d |\delta{\bf r}_k^J|^2, \nonumber \\
\end{eqnarray}
where
\begin{equation}
    \bar{\Delta}_{IJ}^{(d)}=\sqrt{\sum_{k=1}^d \left|\bar{\bf r}_k^I-\bar{\bf r}_k^J\right|^2}
\end{equation}
is the distance measured from the time-averaged configurations.

We assume that the thermal fluctuations are not correlated with the underlying structure. Then, for $d \gg 1$, the cross term
\begin{equation}
    \sum_{k=1}^d (\bar{\bf r}_k^I-\bar{\bf r}_k^J)\cdot(\delta{\bf r}_k^I-\delta{\bf r}_k^J)
\end{equation}
is negligible. We also assume that the fluctuations in the two patches are uncorrelated, so that
\begin{equation}
    \sum_{k=1}^d \delta{\bf r}_k^I \cdot \delta{\bf r}_k^J \approx 0.
\end{equation}
Finally, let $\left[\Delta_T^{(d)} \right]^2$ denote the Debye--Waller contribution~\cite{widmer2006predicting}, or cage size multiplied by $d$, at temperature $T$. We then approximate
\begin{equation}
    \sum_{k=1}^d |\delta{\bf r}_k^I|^2 \simeq \sum_{k=1}^d |\delta{\bf r}_k^J|^2 \simeq \left[\Delta_T^{(d)} \right]^2/2 ,
    \label{eq:DW}
\end{equation}
where the factor of $1/2$ in the last equation comes from accounting for fluctuations at both the initial and final times in the measurement of the Debye--Waller factor.

Under these assumptions, we obtain
\begin{equation}
    \left[\Delta_{IJ}^{(d)}\right]^2 \simeq \left[\bar{\Delta}_{IJ}^{(d)}\right]^2 + \left[\Delta_T^{(d)} \right]^2.
\end{equation}

Since $\epsilon \ge 0$ and $\Delta_{IJ}^{(d)} \ge 0$ by definition, we can write
\begin{eqnarray}
    n_I^{(d)}(\epsilon) &=& \sum_{\substack{J=1\\ (J\neq I)}}^{\mathcal{N}} \theta\!\left( \epsilon - \Delta_{IJ}^{(d)} \right) \nonumber \\
    &=& \sum_{\substack{J=1\\ (J\neq I)}}^{\mathcal{N}} \theta\!\left( \epsilon^2 - \left[ \Delta_{IJ}^{(d)} \right]^2 \right) \nonumber \\
    &\simeq& \sum_{\substack{J=1\\ (J\neq I)}}^{\mathcal{N}} \theta\!\left( {\epsilon'}^{\,2} - \left[ \bar{\Delta}_{IJ}^{(d)} \right]^2 \right),
\end{eqnarray}
where we have introduced the effective threshold $\epsilon'$ defined by
\begin{equation}
    {\epsilon'}^{\,2} = \epsilon^2 - \left[\Delta_T^{(d)}\right]^2 .
\end{equation}

The above result suggests that plotting $C_m^{(d)}$, computed from
instantaneous equilibrium configurations, as a function of $\epsilon'$
may recover a curve similar to that obtained from time-averaged
configurations. Thus, instantaneous configurations alone could, in principle,
be sufficient to extract the scaling regime relevant for the entropy, provided
that the assumptions above are valid and that the cage size is known at each
temperature $T$. Under these conditions, $\left[\Delta_T^{(d)}\right]^2$ can be
estimated as the single-particle cage size multiplied by $d$.

Figure~\ref{fig:thermal_effects}(b) shows the same quantity as in
Fig.~\ref{fig:thermal_effects}(a), but plotted as a function of the effective
threshold $\epsilon'$. Here, we use an independently measured value of
$\left[\Delta_T^{(d)}\right]^2$. The resulting curve is qualitatively similar to those
obtained from thermally averaged and inherent-structure configurations. In
particular, as $d$ increases, the scaling regime becomes progressively more
pronounced. This is encouraging, although it remains difficult to extract the
entropy quantitatively with sufficient precision.

The deviations observed at smaller $d$ may originate from the fact that some of
the assumptions made above are not fully valid in the range of patch sizes
studied here. For example, cross terms may not be negligible for our values of
$d$. In addition, in Eq.~\eqref{eq:DW}, we replace thermal fluctuations by
their mean Debye--Waller value. Since this mean value enters inside the
Heaviside function, a more accurate treatment would be to average over thermal
fluctuations outside the Heaviside function, which effectively acts as a
Gaussian convolution kernel~\cite{kantz2003nonlinear}. Moreover, in two dimensions, the numerical
estimation of $\left[\Delta_T^{(d)}\right]^2$ may be affected by
Mermin--Wagner fluctuations, making it difficult to determine this quantity
precisely~\cite{shiba2016unveiling}. Understanding these different sources of
error, and extending the Grassberger--Procaccia construction to instantaneous
equilibrium configurations, is left for future investigation.

\subsection{Extraction of the Rényi complexities}

From the correlation-integral data $C_m^{(d)}(\epsilon)$, we extract the R\'enyi complexities by two different methods within the scaling regime.
\newline
{\it Intercept-fit method}: Based on the scaling form in Eq.~(\ref{eq:GP_scaling}), this method consists in fitting $\ln C_m^{(d)}(\epsilon)$, for each $d$, to
\[
\ln C_m^{(d)}(\epsilon)=\nu \ln \epsilon + g(d).
\]
We then fit the intercept
\[
g(d)=-d(m-1)s^{\mathrm{R\acute{e}nyi}}_{m} + \ln c
\] 
as a function of $d$ to obtain $s^{\mathrm{R\acute{e}nyi}}_{m}$.

For each function $\ln C_m^{(d)}(\epsilon)$, we determine the most linear subrange $[\epsilon_{\rm min},\epsilon_{\rm max}]$ as the interval over which a linear fit gives the smallest reduced chi-squared. An example analysis script is provided in Ref.~\cite{Gitcode}. This interval $[\epsilon_{\rm min},\epsilon_{\rm max}]$ is also used in the second method described below.

\begin{figure}
\includegraphics[width=\linewidth]{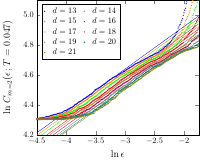}
\includegraphics[width=\linewidth]{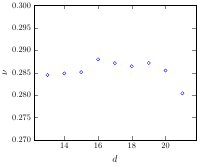}
\includegraphics[width=\linewidth]{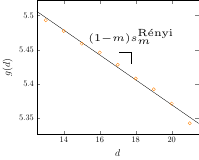}
\caption{Intercept-fit method to extract $s_m^{\rm R\acute{e}nyi}$, illustrated here for $m=2$ and $T=0.047$. Top: For each patch size $d$, the curve $y=\ln C_m^{(d)}$ is fitted to the affine form $y=\nu \ln \epsilon + g(d)$. Center: values of the correlation dimension $\nu$ obtained from these fits, shown for each $d$. Bottom: intercepts $g(d)$ obtained from the fits, shown as a function of $d$. The line is an affine fit of $g(d)$, whose slope yields $(1-m)s_m^{\rm R\acute{e}nyi}$, cf. Eq.~(\ref{eq:GP_scaling}).
%\MO{For (a), the y-axis has log. Is this Log or ln?}\MO{For (b), could you use the y range, [0.27, 0.3]?} \MO{(c) Please use $g(d)$ instead of $b$}
}\label{fig:fits}
\end{figure}

In Fig.~\ref{fig:fits}(a), we show the fitting results for the scaling regime of the correlation integral in the case $m=2$. Figure~\ref{fig:fits}(b) shows that the values of the correlation dimension $\nu$ obtained from these fits depend only weakly on the patch size $d$, which supports the existence of a well-defined scaling regime in $\epsilon$. We find that this weak $d$-dependence becomes more pronounced at higher temperatures, where the scaling regime is less clearly visible by eye. In Fig.~\ref{fig:fits}(c), we show that the intercepts $g(d)$ follow the expected linear behavior as a function of $d$, consistent with Eq.~(\ref{eq:GP_scaling}). From the slope, which is equal to $(1-m)s_m^{\rm R\acute{e}nyi}$, we obtain our estimate of the R\'enyi complexity.
\begin{figure}
\centering
\includegraphics[width=\linewidth]{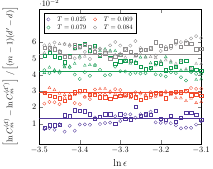}
\caption{Difference between the correlation integrals for patch sizes $d$ and $d'=d+1$, shown for $d=17$ (circles), $d=18$ (squares), and $d=19$ (triangles) for $m=2$. For clarity, other values of $d$ are omitted, and only four temperatures are shown. The horizontal lines indicate the average value, taken over both $\epsilon$ and $d$.
}
\label{fig:diffs_Km}
\end{figure}
\newline\newline
{\it Difference method}: As a second method, from Eq.~(\ref{eq:GP_scaling}), 
we consider the correlation integral for two different patch sizes $d$ and $d' > d$, and compute $s^{\mathrm{R\acute{e}nyi}}_{m}$ as
\begin{equation}
\label{eq:sm_diff}
s^{\mathrm{R\acute{e}nyi}}_{m} 
\simeq 
\frac{
\ln C_m^{(d)}(\epsilon) 
- 
\ln C_m^{(d')}(\epsilon)
}{
(d' - d)(m - 1)
} ,
\end{equation}
where the terms $\ln c+\nu\,\ln\epsilon$ cancel out \cite{grassberger1983estimation}.

The RHS of Eq.~(\ref{eq:sm_diff})
is plotted in Fig.~\ref{fig:diffs_Km} for $m=2$ and four different temperatures. It is seen to be approximately constant over the range of $\epsilon$ corresponding to the scaling regime. To avoid overcrowding the figure, we only show the cases $d=17,18,19$, with $d'=d+1$. The horizontal lines indicate the average over both $\epsilon$ and $d$, and provide an estimate of $s_m^{\rm R\acute{e}nyi}$.

\begin{figure*}
\centering
    \begin{subfigure}{.32\linewidth}
        \includegraphics[width=\linewidth]{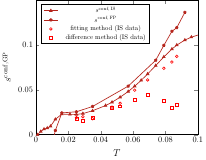}
        \caption{\label{fig:sconf}}
    \end{subfigure}
    \hfill
    \begin{subfigure}{.32\linewidth}
        \includegraphics[width=\linewidth]{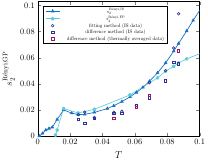}
        \caption{\label{fig:s2}}
    \end{subfigure}
    \hfill
    \begin{subfigure}{.32\linewidth}
        \includegraphics[width=\linewidth]{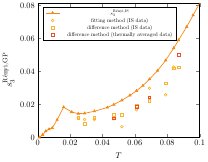}
        \caption{\label{fig:s3}}
    \end{subfigure}
\caption{Configurational entropy ($m\to1$) in panel (a), and R\'enyi complexities with indices $m=2$ in panel (b) and $m=3$ in panel (c), computed using different  approaches. Filled symbols show standard estimates obtained from the inherent-structure distribution (triangles) and from the Franz--Parisi potentials (pentagons). Open symbols show estimates obtained from patch statistics using the intercept-fit method (circles) and the difference method (squares). In addition, for $m=2$, we also show the estimate obtained from thermally averaged data using the difference method (squares). }\label{fig:K123}
\end{figure*}

We report the values of $s_m^{\rm R\acute{e}nyi}$ obtained using the above methods for $m\to1$, $m=2$, and $m=3$ in Fig.~\ref{fig:K123}. We find quantitatively consistent results, especially at sufficiently low
temperatures, where the scaling regime is well identified. At higher
temperatures, however, the estimates obtained from the different methods no
longer agree. This discrepancy arises because the correlation dimension shows
an increasingly pronounced dependence on $d$, so that the
Grassberger--Procaccia scaling assumption in Eq.~\eqref{eq:GP_scaling}, with a
constant $\nu$, ceases to hold. The deviation between the two methods therefore
signals the limit of applicability of the Grassberger--Procaccia method in this
temperature range. We note that the temperature range over which our approach
provides reasonable estimates increases with increasing $m$. This is consistent
with the interpretation that increasing $m$ corresponds to effectively lowering
the temperature~\cite{javerzat2025renyi}.

In Fig.~\ref{fig:K123} we also provide a comparison with estimates obtained from other computational methods for the configurational entropy and the R\'enyi complexities. The estimates extracted from patch-coincidence statistics are in remarkable agreement with those obtained from the probability distribution of inherent structures and from the Franz--Parisi potentials. We further confirm the expected monotonicity of the R\'enyi complexity with respect to $m$, namely
\[
s_{\rm conf} > s_2^{\rm R\acute{e}nyi} > s_3^{\rm R\acute{e}nyi}.
\]

To further support the relevance of the patch method, we also analyzed the correlation integrals computed from thermally averaged configurations for the case $m=2$. The resulting values of the R\'enyi complexity show very good agreement with those obtained from the inherent-structure data.

\section{Intrinsic dimension}
\label{sec:intrinsic_dimension}

We discuss the physical meaning of the exponent $\nu$ extracted from our glass-forming liquid data, and its interpretation in terms of the intrinsic dimensionality of the underlying configurational data~\cite{camastra2016intrinsic}. Because this concept is not commonly used in the glass-physics community, we begin with a brief pedagogical introduction before presenting the results for our glass-forming liquid.

\subsection{Background}

Let us consider $N$ data points scattered in an $M$-dimensional space,
\[
\mathbf{X}_i = \left( X_i^{(1)}, X_i^{(2)}, \ldots, X_i^{(M)} \right)^{\mathrm T} \in \mathbb{R}^M,
\qquad i = 1,2,\ldots,N.
\]
The dimensionality $M$ of the space in which the data are represented is referred to as the \emph{embedding dimension}.

The \emph{intrinsic dimension}, on the other hand, corresponds to the dimensionality of the manifold on which the data are actually distributed or localized. A well-known illustrative example is the Swiss-roll dataset: although the data points are embedded in a three-dimensional space ($M=3$), they lie on a curved two-dimensional surface. In this case, the embedding dimension is $3$, while the intrinsic dimension is $2$.

One way to estimate the intrinsic dimension is through the measurement of the correlation dimension $D_2$, defined via the correlation integral
\begin{equation}
    C(\epsilon) = \frac{1}{N^2} \sum_{\substack{i,j \\ (i \neq j)}}^N \theta\!\left(\epsilon - \lvert \mathbf{X}_i - \mathbf{X}_j \rvert \right),
\end{equation}
where $\theta(\cdot)$ denotes the Heaviside step function.

The correlation dimension $D_2$ is extracted from the small $\epsilon$ scaling behavior of the correlation integral, namely,
\begin{equation}
    C(\epsilon) \sim \epsilon^{D_2}, \qquad \epsilon \to 0.
\end{equation}

It is instructive to consider a few simple examples in an embedding space of dimension $M=3$.
If the $N$ data points are distributed approximately uniformly throughout three-dimensional space, one finds
\(
C(\epsilon) \sim \epsilon^{D_2}
\)
with $D_2 \approx 3$. The same scaling is obtained when the data points are drawn from a three-dimensional Gaussian distribution, since they still fill a volume in $\mathbb{R}^3$.

If, instead, the data are distributed on a two-dimensional manifold embedded in three dimensions, such as in the Swiss-roll example, one observes a correlation dimension $D_2 \approx 2$.
When the data points are aligned along a one-dimensional structure, forming a string-like configuration, the correlation dimension is $D_2 \approx 1$.
In the extreme case where all data points collapse onto a single point, one finds $D_2 \approx 0$. The same value is obtained when the data are localized on a finite number of isolated points.

These examples illustrate that the correlation dimension $D_2$ characterizes how data points are distributed and organized in the embedding space, and provides a quantitative measure of their effective, or intrinsic, dimensionality.

\subsection{Glass-forming liquid data}
\begin{figure}[!ht]    
\includegraphics[width=\linewidth]{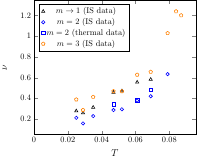}
    \caption{Temperature dependence of the exponent $\nu$ for $m=2$, $3$, and in the limit $m \to 1$, obtained from fits of the correlation integral $C_m^{(d)}$.}
    \label{fig:nu}
\end{figure}

The original embedding dimensionality of our data is $2d$, corresponding to $d$ particles in two spatial dimensions. 
The value of the correlation dimension $\nu$ as a function of temperature is shown in Fig.~\ref{fig:nu}, for $m=2$, $m=3$, and in the limit $m \to 1$. For each temperature, $\nu$ is obtained by averaging over $d$ the slopes extracted from fits of the correlation integral $C_m^{(d)}$. Although our estimates of $\nu$ are not very precise, they clearly show that $\nu$ decreases monotonically as the temperature is lowered and, in this regime, appears to be independent of $m$ and of the type of data considered.
% We also find that the extracted values of $\nu$ are significantly smaller than $2d$, indicating that patch configurations populate the data (or phase) space in a highly organized and constrained manner. In particular, $\nu$ is of order unity and decreases with decreasing temperature. This behavior suggests that patch configurations become increasingly localized in phase space as $T \to 0$, forming mutually disconnected regions. 
% This phenomenology is reminiscent of the mean-field picture of the glass transition and related scenarios, as illustrated, for instance, in Fig.~4 of Ref.~\cite{parisi2010mean} and Fig.~2 of Ref.~\cite{krzakala2007gibbs}. At the same time, our numerical analysis provides a novel and quantitative way to probe the organization of phase space in a finite-dimensional glass-forming system.

%The scaling form proposed by Grassberger and Procaccia in Eq.~(\ref{eq:GP_scaling}) yields the exponent $\nu$, which, for $m=2$, can be identified with the correlation dimension.

%\newpage
%\bibliography{reference}% Produces the bibliography via BibTeX.

\end{document}